# Simulations of the inner magnetospheric energetic electrons using the IMPTAM-VERB coupled model

Angelica M. Castillo [a,b,*], Yuri Y. Shprits [a,b,c], Natalia Ganushkina [d,e], Alexander Drozdov [c], Nikita Aseev [a,b], Dedong Wang [a], Stepan Dubyagin [e]

[a] GFZ German Research Centre For Geosciences, Potsdam, Germany
[b] Institute of Physics and Astronomy, University of Potsdam, Potsdam, Germany
[c] Department of Earth, Planetary and Space Sciences, University of California, Los Angeles, CA, USA
[d] Climate and Space Sciences and Engineering, University of Michigan, Ann Arbor, MI, USA
[e] Finnish Meteorological Institute, Helsinki, Finland

## ARTICLE INFO

## ABSTRACT

In this study, we present initial results of the coupling between the Inner Magnetospheric Particle Transport and Acceleration Model (IMPTAM) and the Versatile Electron Radiation Belt (VERB-3D) code. IMPTAM traces electrons of $10-100$ keV energies from the plasma sheet ($L = 9$ Re) to inner L-shell regions. The flux evolution modeled by IMPTAM is used at the low energy and outer $L^*$ computational boundaries of the VERB code (assuming a dipole approximation) to perform radiation belt simulations of energetic electrons. The model was tested on the March 17th, 2013 storm, for a six-day period. Four different simulations were performed and their results compared to satellites observations from Van Allen probes and GOES. The coupled IMPTAM-VERB model reproduces evolution and storm-time features of electron fluxes throughout the studied storm in agreement with the satellite data (within $\sim 0.5$ orders of magnitude). Including dynamics of the low energy population at $L^* = 6.6$ increases fluxes closer to the heart of the belt and has a strong impact in the VERB simulations at all energies. However, inclusion of magnetopause losses leads to drastic flux decreases even below $L^* = 3$. The dynamics of low energy electrons (max. 10s of keV) do not affect electron fluxes at energies $\geq 900$ keV. Since the IMPTAM-VERB coupled model is only driven by solar wind parameters and the Dst and Kp indexes, it is suitable as a forecasting tool. In this study, we demonstrate that the estimation of electron dynamics with satellite-data-independent models is possible and very accurate.

## 1. Introduction

Highly energetic electrons (100s of keV to a few MeV) trapped in the Van Allen belts can produce surface or deep dielectric charging on spacecraft materials and damage their electronics (Baker et al., 1996). Strong geomagnetic storms enhance the risk of such operational failures in satellites at all orbits, specially those located in the radiation regions (Lanzerotti, 2001; Odenwald et al., 2006). While the inner radiation belt shows a rather stable behavior, the outer belt is highly variable (Rothwell and McIlwain, 1960; Craven, 1966) and its dynamics are strongly dependent on geomagnetic activity.

Inside geostationary orbit, radiation belt electrons are accelerated and lost by a number of processes occurring throughout the inner magnetospheric environment (Shprits et al., 2008a,b; Millan and Baker, 2012). Once the particles in the magnetotail convect into the radiation belts (Baker and Stone, 1978; Elkington et al., 2004), their interaction with a variety of plasma waves (e.g. ULF, VLF, ELF waves) will also determine the course of their evolution. Under the assumption that collisionless charged particles in the ambient magnetic field experience resonant interactions with incoherent electromagnetic waves whose amplitudes are rather small (Kennel and Engelmann, 1966; Lerche, 1968; Schulz and Lanzerotti, 1974), the Fokker–Planck diffusion equation derived from quasi-linear theory describes the violation of the adiabatic invariants of particle motion caused by the processes/interactions mentioned above and the thereby resulting evolution of electron Phase Space Density (**PSD** or $f$) in terms of radial distance, energy and pitch angle (Schulz and Lanzerotti, 1974; Shprits et al., 2008a). The quasi-linear diffusion rates needed to solve the equation can be estimated using a high plasma density approximation (Lyons et al., 1971) or using alternative methods without this approximation (e.g. Glauert and Horne, 2005; Albert and Young, 2005; Albert, 2007; Ni et al., 2008; Sprits and Ni, 2009).

Modeling studies accounting for diffusion due to chorus waves showed that MeV electrons can be efficiently scattered and accelerated during resonant wave-particle interactions in the radiation belts (Varotsou et al., 2005; Horne et al., 2005), and that the interplay of radial diffusion and local acceleration can lead to the acceleration of keV electrons to MeV energies in the outer belt (Varotsou et al., 2008; Shprits et al., 2009b). Particles injected into the inner magnetosphere can undergo acceleration by chorus waves, which results in peaks in phase space density (or electron flux) around $L^* = 4$–$5$. The importance of enhanced particle losses by outward radial diffusion, after the peak is formed, has also been appointed (Reeves et al., 1998; Brautigam and Albert, 2000; Miyoshi et al., 2003, 2006). Depletion of energetic fluxes observed at different energies for certain geomagnetic events, indicated the fundamental role of magnetopause losses and outward diffusion for the dynamics of the radiation belts (Shprits et al., 2006a). Magnetospheric convection is the dominant transport mechanism of injected low energy electrons, while radial diffusion is particularly important for the inward transport of high energy particles, e.g Jordanova and Miyoshi (2005), Chen et al. (2015), Shprits et al. (2015).

Electron fluxes of different energy populations in the inner magnetosphere vary by several orders of magnitude during geomagnetic storms. In particular, it is considered that low energy electrons (10s of keV) play a key role in the generation and amplification of lower band chorus waves, which in turn, can effectively accelerate substorm injected electrons (up to a few 100s of keV) even up to multi-MeV energies in the outer belt during high levels of geomagnetic activity (Lyons, 1984; Horne and Thorne, 1998; Summers et al., 1998; Horne et al., 2007; Jaynes et al., 2015). To study the coupling of different particle populations in the inner magnetosphere under different geomagnetic conditions, a few convection–diffusion models have been developed in the past 20 years. The kinetic Ring current-Atmosphere interactions Model (RAM) (Jordanova et al., 1997) solves the bounce-averaged kinetic equation for the phase space distribution function of the main inner magnetospheric species, including multi-MeV electrons. The modified RAM code accounts for convection, radial diffusion and for losses due to: charge exchange, Coulomb collisions, wave particle interactions and precipitation to the atmosphere (Jordanova and Miyoshi, 2005). The Radiation Belt Environment (RBE) model (Fok et al., 2008) solves the bounced-averaged Boltzmann transport equation for the distribution function of energetic electrons. RBE accounts for particle drifts along magnetic field lines, variable magnetospheric convection, losses to the loss cone, wave particle interactions with chorus waves, pitch-angle-, energy- and cross-diffusion. RBE does not include an explicit radial diffusion term (Fok et al., 2011). The VERB-4D code was developed based on the 1D, 2D and 3D VERB-codes (Shprits et al., 2005, 2006b, 2008b, 2009b; Subbotin and Shprits, 2009; Subbotin et al., 2010; Kim et al., 2011; Subbotin et al., 2011a,b). VERB-4D solves the modified Fokker–Planck diffusion equation with additional convection terms to obtain the evolution of electron PSD (Subbotin and Shprits, 2012; Shprits et al., 2015). This model accounts for radial-, pitch angle-, energy- and cross-diffusion, wave particle interactions with large number of waves, losses to the atmosphere, losses due to pitch angle scattering and losses to the magnetopause.

Although, some advective–diffusive codes ignore local acceleration or radial diffusion and focus on low energy particles, they include more realistic simulation conditions and thereby high accuracy. The dependence between parameters in such models is often difficult to assess due to their numerical complexity and high dimensionality, which also makes the simulations computationally very expensive and disk space demanding, even for storm specific simulations. On the other hand, process focused (only-convection or -diffusion) models have simpler frameworks that allow for less computationally expensive simulations, reasonable disk space requirements and short running times for long-term simulations, e.g. VERB-3D performs a one year simulation in ~ 4.5 hours with a time step of one hour. For these reasons, process focused codes are the most appropriate tools to study and understand long-term dynamics of magnetospheric plasmas.

Many modeling radiation belt studies assumed a constant low energy boundary (e.g. Subbotin et al., 2011a,b; Drozdov et al., 2015) or obtained it from observations inside geosynchronous orbit (e.g. Albert et al., 2009; Tu et al., 2014; Glauert et al., 2014; Li et al., 2014), and estimated the outer radial boundary from satellite data at $L = 5.5$ Re or $L = 6.6$ Re, which strongly drives the simulations due its proximity to the heart of the belt. Using satellite data at the boundaries of diffusion codes provides accurate simulation results, however, this method does not allow us to disentangle the dynamics of different electron populations and is not suitable for forecasting purposes. One-way coupling of process focused models is an alternative approach to estimate diffusion boundary conditions and to avoid the computational disadvantages of advective–diffusive codes while reaching fairly accurate simulation conditions and results. Through model coupling, we can also study the role of particular electron populations in the physical mechanisms controlling acceleration and loss in the inner magnetosphere.

In the current study, we coupled the Inner Magnetospheric Particle Transport and Acceleration Model (IMPTAM) (Ganushkina et al., 2013, 2014, 2015) with the three-dimensional Versatile Electron Radiation Belt code (VERB-3D) (Subbotin and Shprits, 2009; Shprits et al., 2009b) to obtain a diffusion model of the Earth's radiation belts that accounts for the dynamics of the 10s of keV electron population in order to model the evolution of energetic electrons ($\geq 400$ keV). The flux evolution of low energy electrons in the plasma sheet (outer boundary at $L = 9$ Re) and inside geosynchronous orbit is computed with the convection model IMPTAM and then used as the input boundary conditions of the diffusion code VERB-3D. In the past, Subbotin et al. (2011a) coupled VERB-3D with the Rice Convection Model (RCM) to model the April 21, 2001 geomagnetic storm. RCM computes magnetospheric electric fields self-consistently and calculates drifts assuming isotropic pitch angle distributions (Toffoletto et al., 2003). Unlike RCM, IMPTAM accounts for radial diffusion and pitch-angle scattering due to wave-particle interactions with hiss and chorus waves. Furthermore, while the outer boundary of the RCM code was obtained from extrapolated Geotail data, IMPTAM is only driven by time dependent solar wind and Interplanetary Magnetic Field (IMF) parameters, making our approach satellite data independent. VERB-3D has also been upgraded over the past few years, now including mixed diffusion, magnetopause losses and improved hiss diffusion coefficients. Additionally, we present quantitative and qualitative comparisons with satellite observations from 4 different spacecrafts. Since, the IMPTAM-VERB coupled model does not use satellite data based inputs, it is suitable as a forecasting tool.

The numerical models, wave models, boundary and initial conditions underlying the IMPTAM and VERB codes are described in detail in Section 2. The March 17th, 2013 event and relevant event specific studies are presented in Section 3. The results of the IMPTAM simulation used as boundary conditions for VERB-3D are described in Section 4, together with the data processing performed, in order to make them suitable for the coupling process. Section 5 presents the results of four different simulations. The summary and conclusions of this study are given in Section 6.

## 2. Coupling strategy and model description

In spite of the complexity of electron dynamics in the Earth's magnetosphere, with the following key points, we can simplify the modeling and coupling approach: (1) Low energy electrons (up to few 100s of keV) injected in the plasma sheet during geomagnetic events are transported earthward by means of convection and during substorm associated dipolarization events (Baker et al., 1996; Birn et al., 1997; Fu et al., 2011), they experience $E \times B$-drift and their motion follows the conservation of the first adiabatic invariant ($\mu$), which leads to their energization as they move towards the Earth (Jordanova and Miyoshi, 2005; Liu et al., 2005); (2) Electrons of $10 - 100$ keV energies at $L^* \approx 5 - 7$ Re are equally affected by diffusive and convective processes (Subbotin et al., 2011a; Shprits et al., 2015); (3) while the storm-time enhancement of low energy electrons ($10 - 50$ keV) at $L^* \approx 3 - 5$ Re can be explained by convection only, electrons of energies $\geq 100$ keV at the same radial distances are mostly affected by diffusion (Liu et al., 2003) and are strongly subject to gradient and curvature drifts.

In the present study, we assume that electrons at $10 - 100$ keV at $L^* \leq 6.6$ are equally affected by diffusive and convective processes. With this approach, we can use one-way code coupling to combine available physics based models that calculate the electron evolution of two different electron populations: (a) energies of $10–100$ keV, modeled with IMPTAM and (b) energies $\geq 100$ keV modeled by VERB.

## 2.1. Modeling low energy electrons with IMPTAM

IMPTAM (Ganushkina et al., 2013, 2014, 2015) traces distributions of electrons in the drift approximation with arbitrary pitch angles from the plasma sheet to the inner L-shell regions with energies reaching up to hundreds of keV in time dependent magnetic and electric fields. The distribution of particles is traced in the drift approximation taking into account the $E \times B$ drift and the magnetic drifts with bounce-averaged drift velocities. Relativistic effects in the drift velocities of electrons are also considered.

To follow the evolution of the particle distribution function $F$ and particle fluxes in the inner magnetosphere dependent on the position $R$, time $t$, kinetic energy $E_{kin}$, and pitch angle $\alpha$, it is necessary to specify certain parameters: (1) the particle distribution at initial time at the model boundary; (2) time dependent magnetic and electric fields at all locations; (3) the drift velocities; (4) all sources and losses of particles. The changes in the distribution function $F(R, \varphi, t, E_{kin}, \alpha)$ are obtained by solving the equation:

$$\frac{\partial F}{\partial t} = \frac{\partial F}{\partial \varphi} V_\varphi + \frac{\partial F}{\partial R} V_R + sources - losses, \tag{1}$$

where $R$ and $\varphi$ are the radial and azimuthal coordinates in the equatorial plane, respectively, $V_\varphi$ and $V_R$ are the azimuthal and radial components of the bounce-averaged drift velocity. The model boundary can be set in the plasma sheet at distances, depending on the scientific questions we are trying to answer, from $6.6\ R_E$ to $10\ R_E$. Liouville's theorem is used to gain information about the entire distribution function by mapping the boundary conditions throughout the simulation domain, including loss process attenuation, through the time-varying magnetic and electric fields. Potential electron energization is accounted for by solving the radial diffusion equation (Schulz and Lanzerotti, 1974), Eq. (2), for the obtained distribution function:

$$L^2 \frac{\partial}{\partial L} \left( \frac{1}{L^2} D_{LL} \frac{\partial f}{\partial L} \right) - \frac{f}{\tau} = \frac{\partial f}{\partial t}. \tag{2}$$

where $L$ is the radial distance, $\tau$ is the electron lifetime and $f$ is the electron phase space density. Different approaches for the estimation of the radial diffusion coefficient $D_{LL}$ have been presented in the past. Brautigam and Albert (2000) determined the values of the electromagnetic and electrostatic components, $D_{LL}^{EM}$ and $D_{LL}^{ES}$ (respectively), using empirical Kp parametrizations. Fei et al. (2006) developed a method for the estimation of event specific time-dependent $D_{LL}$-coefficients using the ULF electric and magnetic field power spectral density obtained from computationally expensive magnetohydrodynamics (MHD) simulations (e.g. Ozeke et al., 2014; Ali et al., 2016). Li et al. (2016, 2017) calculated radial diffusion coefficients for the recovery phase of the March 17, 2013 storm using Fei et al. (2006) formalism. Also, Su et al. (2015) presented a data-driven method for the calculation of $D_{LL}$ that uses measurements of the power spectral density.

In order to validate the coupled model and make it suitable for forecasting purposes, general and easily implementable estimation methods for $D_{LL}$, s.a. statistical or parametrized models, are needed. Drozdov et al. (2017) compared the performance of $D_{LL}$ coefficients of Brautigam and Albert (2000) and Ozeke et al. (2014) showing that both models deliver almost identical 3D simulation results. Therefore, for this study, we chose the method of Brautigam and Albert (2000). For this parametrization, diffusion due to magnetic field fluctuations at $L > 3$ dominates over the diffusion produced by electrostatic field fluctuations (Kim et al., 2011), therefore the electrostatic component of the radial diffusion coefficient is ignored here. The Kp-dependent magnetic component of the radial diffusion coefficients $D_{LL}$ with units [day$^{-1}$] (Eq. (3)) are computed following Brautigam and Albert (2000).

$$D_{LL} = 10^{0.506K_p - 9.325} L^{10} \tag{3}$$

The computations described above are repeated in the same order for each step of the simulation. The model accounts for convective outflow, Coulomb collisions and loss to the atmosphere. We assume strong pitch angle scattering at the distances where the ratio ($R_c / \rho$) between the radius of the field line curvature in the equatorial current sheet ($R_c$) and the effective Larmor radius $\rho$ is between $6 - 10$ (Sergeev and Tsyganenko, 1982; Büchner and Zelenyi, 1987; Delcourt et al., 1996). Precipitation to the atmosphere is calculated following Jordanova et al. (2008), with a time scale of a quarter bounce period, and the loss cone corresponds to an altitude of 200 km. Electron losses caused by pitch angle diffusion due to wave-particle interactions are also included in the simulations using the parameterizations of the electron lifetimes due to interactions with chorus and hiss waves given by Orlova and Shprits (2014) and Orlova et al. (2014, 2016) with the activity dependent plasmapause location by Carpenter and Anderson (1992):

$$L_{pp} = 5.6 - 0.46 K p_{max24} \tag{4}$$

where $K p_{max24}$ is the maximum value of Kp in the previous 24 h. The use of a symmetric plasmapause location is justified, because VERB-3D requires the MLT-averaging of the boundary condition fluxes computed by IMPTAM. In general, IMPTAM is designed to perform simulations using any given magnetic or electric field model, including a self-consistent magnetic field. In addition to the large-scale fields, transient fields associated with dipolarization processes in the magnetotail during substorm onset can be modeled by IMPTAM (e.g., Ganushkina et al. (2005, 2013, 2014)) as an earthward propagating electromagnetic pulse of localized radial and longitudinal extent (Li et al., 1998; Sarris et al., 2002). In this study, we do not take these fields into account, since we do not resolve for substorm dynamics and, therefore, consider their effects to be not significant. We focus on low energy electrons and assume that their contribution to the total magnetic pressure is negligible, because their contribution to the ring current is less than 10% and thus they generate an even smaller distortion of the background field.

### 2.1.1. Setup of IMPTAM simulations

IMPTAM is driven by various solar wind (SW) and interplanetary magnetic field (IMF) parameters, as well as different geomagnetic indices. We used Tsyganenko T96 magnetic field model (Tsyganenko, 1995), which uses the Dst index, the dynamic pressure of the solar wind ($P_{dyn}$) and the y and z components of the IMF ($B_Y$ and $B_Z$) as input parameters. The electric field was determined using the mean velocity of the solar wind ($V_{SW}$), the strength of the IMF ($B_{IMF}$), $B_Y$ and $B_Z$ (via IMF clock angle $\theta_{IMF}$) by mapping the Boyle et al. (1997) ionospheric potential $\Phi$ to the magnetosphere. The model boundary is set at $9\ R_E$ and a kappa electron distribution function ($\kappa = 1.5$) is used, which has the best agreement with electron fluxes at $50–150$ keV energies observed by the LANL satellites at geostationary orbit (Horne et al., 2013). We assume that the distribution can be fitted by the kappa shape only in a finite range of velocities. The parameters of the kappa distribution function are the number density $N_{ps}$ and temperature $T_{ps}$ in the plasma sheet given by the Dubyagin et al. (2016) empirical model at distances between 6 and 11 $R_E$, based on THEMIS data. The number density in the plasma sheet ($N_{ps}$) is driven by the solar wind number density ($N_{SW}$) and the southward component of the IMF ($B_S$). The temperature in the plasma sheet ($T_{ps}$) is dependent on $V_{sw}$, the southward ($B_S$) and northward ($B_N$) components of the IMF.

## 2.2. Modeling the radiation belts with VERB-3D

The modified Fokker–Planck equation describes time-changes of the phase-averaged phase space density (PSD or $f$) in the magnetosphere of the Earth, in terms of the three adiabatic invariants (Schulz and Lanzerotti, 1974; Walt, 1994). Using bounce and drift averaged diffusion coefficients ($D_{L^*L^*}$, $D_{pp}$, $D_{p\alpha_0}$, $D_{\alpha_0\alpha_0}$), this equation can be transformed into ($L^*$, $p$, $\alpha_0$) coordinates and is known as the bounce- and drift-averaged Fokker–Planck-equation (Shprits et al., 2008b; Subbotin et al., 2010):

$$
\begin{aligned}
\frac{\partial f}{\partial t} &= L^{*2} \frac{\partial}{\partial L^*}\bigg|_{\mu,J} \left( \frac{1}{L^{*2}} D_{L^*L^*} \frac{\partial f}{\partial L^*}\bigg|_{\mu,J} \right) + \\
&\quad \frac{1}{p^2} \frac{\partial}{\partial p}\bigg|_{\alpha_0,L^*} p^2 \left( D_{pp} \frac{\partial f}{\partial p}\bigg|_{\alpha_0,L^*} + D_{p\alpha_0} \frac{\partial f}{\partial \alpha_0}\bigg|_{p,L^*} \right) + \\
&\quad \frac{1}{T(\alpha_0)\sin(2\alpha_0)} \frac{\partial}{\partial \alpha_0}\bigg|_{p,L^*} T(\alpha_0)\sin(2\alpha_0)\times \\
&\quad \left( D_{\alpha_0\alpha_0} \frac{\partial f}{\partial \alpha_0}\bigg|_{p,L^*} + D_{\alpha_0 p} \frac{\partial f}{\partial p}\bigg|_{\alpha_0,L^*} \right) - \frac{f}{\tau_{lc}},
\end{aligned}
$$

(5)

where $\alpha_0$ is the equatorial pitch angle, $p$ is the relativistic momentum and $L^* = (2\pi M)/(\Phi R_E)$, with $M$ the magnetic moment. $T(\alpha_0)$ is an approximation of the bounce frequency in a dipole field and is estimated after Lenchek et al. (1961) as:

$$
T(\alpha_0) = 1.3802 - 0.3198 \left( \sin\alpha_0 + \sin^{1/2}\alpha_0 \right)
$$

(6)

Non-adiabatic particle motion caused, for example, by rapid electromagnetic fluctuations can violate the conservation of some adiabatic invariants and lead to transport by diffusion. In Eq. (5) the radial diffusion of particles in terms of PSD is described by the first term on the right hand side, where $D_{L^*L^*}$ is the radial diffusion coefficient. In contrast to the other terms of the equation, the radial diffusion term is written in terms of $L^*, \mu, J$, which is necessary as radial diffusion leads to particle transport along constant $\mu$ and $J$, but does not conserve $L^*$ (Schulz and Lanzerotti, 1974). Also, adiabatic motion of particles due to slow variations in the magnetic field configuration occurs under conservation of all three adiabatic invariants and can be accounted for by using these phase space coordinates.

The second, third and fourth terms on the right hand side of Eq. (5) describe local processes. Momentum diffusion with diffusion coefficient $D_{pp}$ is given by the second term. The pitch angle diffusion process is described by the third term, where $D_{\alpha_0\alpha_0}$ is the diffusion coefficient, and the fourth term estimates dynamics due to mixed diffusion, where $D_{p\alpha_0} = D_{\alpha_0 p}$ is the corresponding diffusion coefficient. The last term on the right hand side of Eq. (5), ($f/\tau_{lc}$), accounts for the losses inside the loss cone. Here, $\tau_{lc}$ is a characteristic lifetime assumed to be infinite for particles with pitch angles outside the loss cone and otherwise, equal to quarter of a bounce period.

### 2.2.1. Diffusion coefficients

The magnetic component of the radial diffusion coefficient ($D_{L^*L^*}$) is calculated using Eq. (3) (Brautigam and Albert, 2000) for $L^*$ and used by the VERB-code for all $Kp$ values. Following Kim et al. (2011), the electric component of the radial diffusion coefficient is not taken into account in the VERB-3D simulations. VERB-3D simulations account for wave particle interactions with day and night side chorus waves, plasmaspheric hiss waves, lightning-generated whistler waves and anthropogenic VLF waves. Table 1 presents an overview of the wave models used for the estimation of diffusion coefficients. For a simplified computation of the local scattering rates, a dipole geometry was adopted. Day and night side chorus wave parametrizations are taken from Li et al. (2007). Chorus waves dominate the diffusion outside the plasmasphere. However, wave properties have different effects on particle diffusion and are distributed differently in the day and night sectors. Day side chorus waves scatter electrons in the outer radiation belt into the loss cone causing rapid particle losses, while night side chorus is mostly responsible for the acceleration of particles observed during the recovery phase of geomagnetic storms (Li et al., 2007). The diffusion coefficients due to chorus waves were computed with the Full Diffusion Code (FDC) (Ni et al., 2008; Shprits and Ni, 2009) for resonance orders of up to five.

Hiss waves, lightning-generated whistler waves and VLF waves of anthropogenic origin produce particle losses inside the plasmasphere. For the parametrization of plasmaspheric hiss, the diffusion coefficients from Orlova et al. (2014) were used. This model is based on quadratic fits to the hiss amplitudes on day and night side, as a function of L-shell, $Kp$ and latitude. The fits were estimated using data from the CRRES wave experiment, considering the increase in obliquity of hiss waves as these propagate along the field line (Agapitov et al., 2013; Thorne et al., 2013). Similar to previous studies, the frequency spectrum of hiss waves is approximated by the Gaussian-function. Wave parameters for lightning-generated waves are calculated following Meredith et al. (2007), while parameters for anthropogenic VLF waves are based on Abel and Thorne (1998), and Starks et al. (2008). Diffusion rates estimated for chorus, hiss and lightning-generated whistler waves were averaged over magnetic local time (MLT) and scaled throughout the simulation using the $Kp$-index and the variation of wave power, in order to represent changes in wave activity during storm-time (Carpenter and Anderson, 1992; Sheeley et al., 2001; Shprits et al., 2007b). The plasmapause position is calculated after Carpenter and Anderson (1992) (Eq. (4)).

### 2.2.2. Boundary and initial conditions

The VERB-code computes the numerical solution of Eq. (5) using a fully implicit finite differences method on a high resolution grid of ($46\times 101\times 91$) points for radial distance, energy and equatorial pitch angle, respectively (Drozdov et al., 2015). In order to obtain better resolution in high-PSD regions, as observed at low energies and at the edge of the loss cone, logarithmic distribution is used for energy and equatorial pitch angle grid points (Subbotin et al., 2011b). The contribution of radial, local and mixed diffusion processes to the total PSD variation are calculated as a single implicit operator, which provides high stability to the code (Subbotin et al., 2010).

The bounce-averaged Fokker–Planck equation (Eq. (5)) is solved for a range of $L^*$ from 1 to 6.6 and for equatorial pitch angles from $0.7°$ to $89.3°$. Selecting $L^* = 6.6$ Re as the outer boundary of the radial diffusion operator is reasonable, since this is commonly a closed drift shell and physics of radial diffusion apply for particles inside geosynchronous orbit (Subbotin et al., 2011b). In order to estimate the lower energy boundary, we have to take into account that electrons moving earthward conserve the first and second adiabatic invariants ($\mu$ and J, respectively), and undergo energization due to the increasing magnetic field strength (Schulz and Lanzerotti, 1974). Also, the low energy boundary should not be chosen below 10 keV, as the dynamics of particles at lower energies are rather less influenced by diffusion processes (Liu et al., 2003; Horne et al., 2005). Choosing $\mu = 9.3634$ MeV/G as the low energy boundary allows us to resolve energies around 10 keV at the outer radial boundary ($L^* = 6.6$), energies of ~80 keV around $L^* = 4$ and about 1.3 MeV at $L^* = 1$. The computational grid of VERB-3D is irregular in the energy range (see Figure 5 in (Subbotin et al., 2011a), i.e. energy values increase with decreasing $L^*$ (Subbotin and Shprits, 2009).

For each operator (radial distance, energy and pitch angle), two boundary conditions, one upper and one lower PSD value, are needed in order to perform a VERB simulation. Table 2 presents a summary of the boundary conditions used in this study. A zero PSD-derivative at the upper $\alpha$-boundary ($\alpha_0 = 89.3°$) describes a flat pitch angle distribution (Shprits et al., 2007a). Setting the PSD-derivative equal to zero at the lower pitch angle boundary ($\alpha_0 = 0.7°$), we account for strong diffusion regimes (Shprits et al., 2009a) and reduce possible negative values generated by fluxes crossing the boundary (Albert, 2013). At the inner boundary ($L^* = 1$), PSD is equal to zero. At the upper energy boundary, a zero PSD boundary condition is applied, representing the absence of high-energy particles (>10 MeV) electrons (Subbotin and Shprits, 2009; Subbotin et al., 2011b). Initial PSD values are calculated as the steady state solution of the radial diffusion equation Eq. (2) by estimating the diffusion coefficient $D_{L^*L^*}$ for quite times (Kp = 2) and setting the electron lifetime $\tau$ equal to three days (Shprits et al., 2005). In this study, time dependent variations of the lower energy and upper $L^*$ boundaries are modeled using IMPTAM.

**Table 1**
Wave parameters used for the computation of diffusion coefficients.

| Wave type | $B_w$ (pT) | $\lambda_{max}$ | Density model | Percent MLT | Wave spectral properties | Distribution in wave normal |
|---|---|---|---|---|---|---|
| Chorus day | $10^{0.75+0.04\lambda}(2 \times 10^{0.73+0.91Kp})^{0.5}/57.6$ for $Kp \leq 2+$; $10^{0.75+0.04\lambda}(2 \times 10^{2.5+0.18Kp})^{0.5}/57.6$ for $2+ < Kp \leq 6$ | 35° | Sheeley et al. (2001) | 25% | $\omega_m/\Omega_e = 0.2$, $\delta\omega/\Omega_e = 0.1$, $\omega_{uc}/\Omega_e = 0.3$, $\omega_{lc}/\Omega_e = 0.1$. | $\theta_m = 0°$, $\delta\theta = 30°$, $\theta_{uc} = 45°$, $\theta_{lc} = 0°$ |
| Chorus night | $50(2 \times 10^{0.73+0.91Kp})^{0.5}/57.6$ for $Kp \leq 2+$; $50(2 \times 10^{2.5+0.18Kp})^{0.5}/57.6$ for $2+ < Kp \leq 6$ | 15° | Sheeley et al. (2001) | 25% | $\omega_m/\Omega_e = 0.35$, $\delta\omega/\Omega_e = 0.15$, $\omega_{uc}/\Omega_e = 0.65$, $\omega_{lc}/\Omega_e = 0.65$. | $\theta_m = 0°$, $\delta\theta = 30°$, $\theta_{uc} = 45°$, $\theta_{lc} = 0°$ |
| Plasmaspheric Hiss | See Orlova et al. (2014) | 45° | Denton et al. (2006) | 62.5% | $f_m = 550$ Hz, $\delta f = 300$ Hz, $f_{uc} = 2000$ Hz, $f_{lc} = 100$ Hz | $\theta_m = 0°$, $\delta\theta = 30°$, $\theta_{uc} = 45°$, $\theta_{lc} = 0°$ |
| Lightning generated whistlers | 3 | 45° | Carpenter and Anderson (1992) | 100% | $f_m = 2000$ Hz, $\delta f = 4500$ Hz, $f_{uc} = 6500$ Hz, $f_{lc} = 2500$ Hz | $\theta_m = 0°$, $\delta\theta = 22.5°$, $\theta_{uc} = 22.5°$, $\theta_{lc} = 67.5°$ |
| Anthropogenic 1 | 0.8 | 45° | Carpenter and Anderson (1992) | $4 \times 2.4\%$ | $f_m = 17100$ Hz, $\delta f = 50$ Hz, $f_{uc} = 17000$ Hz, $f_{lc} = 17200$ Hz | $\theta_m = 45°$, $\delta\theta = 22.5°$, $\theta_{uc} = 22.5°$, $\theta_{lc} = 67.5°$ |
| Anthropogenic 2 | 0.8 | 45° | Carpenter and Anderson (1992) | $4 \times 2.4\%$ | $f_m = 22300$ Hz, $\delta f = 50$ Hz, $f_{uc} = 22400$ Hz, $f_{lc} = 22200$ Hz | $\theta_m = 45°$, $\delta\theta = 22.5°$, $\theta_{uc} = 22.5°$, $\theta_{lc} = 67.5°$ |

**Table 2**
Boundary conditions used for the IMPTAM-VERB coupled simulations. The blue highlight shows the time dependent boundary conditions taken from the IMPTAM computations.

| Boundary | Condition | Underlying physical processes |
|---|---|---|
| $\alpha_0 = 0.7°$ | $\partial(PSD)/\partial\alpha_0 = 0$ | Strong diffusion regimes |
| $\alpha_0 = 89.3°$ | $\partial(PSD)/\partial\alpha_0 = 0$ | Flat pitch angle distribution |
| $L^* = 1$ | $PSD = 0$ | Losses to the atmosphere |
| $L^* = 6.6$ | $PSD(time)$ | Coupling with IMPTAM |
| $E = E_{min}$ | $PSD(time)$ | Coupling with IMPTAM |
| $E = E_{max}$ | $PSD = 0$ | Absence of multi-MeV energy electrons |

## 3. March 17th, 2013 storm

For the initial IMPTAM-VERB coupled simulations, we have chosen to study the March 17th, 2013 geomagnetic storm, which has received much attention in the scientific community, as it has been one of the strongest geomagnetic events during the lifetime of the Van Allen Probe mission. Many event specific studies have been published. Yu et al. (2014) used a two-way coupled model between RAM (Jordanova et al., 2012) and a global MHD model (BATS-R-US-code) (Zaharia et al., 2010) to simulate substorm injections observed by the instruments on board Van Allen Probes during this event. Their study focuses on low energetic particles and orbital observations over 24 h. Their results suggest that impulsive source plasma sheet injections and a large convection electric field are necessary to develop the observed strong ring current. Boyd et al. (2014) quantified the storm specific dependence of PSD gradients on the adiabatic invariant $\mu$ based on Magnetic Electron Ion Spectrometer (MagEIS) data. They observed that the injection of seed electrons (a few 100s of keV) from the plasma sheet is followed by local acceleration of up several MeV energies. Li et al. (2014) studied electron acceleration due to particle interactions with chorus waves during this geomagnetic event and built a global distribution model of chorus wave intensity using measurements of multiple Polar Orbiting Environment Satellites (POES). Electron dynamics simulated with this wave model suggests that local acceleration by chorus waves plays a key role in accelerating injected seed electrons ($\approx$100 keV) to multi-MeV energies. Ripoll et al. (2017) calculated time dependent diffusion coefficients and loss rates from hiss wave properties and ambient plasma data measured by the Electric and Magnetic Field Instrument Suite and Integrated Science (EMFISIS) during March 2013. Their 1-D simulations show that hiss wave activity combined with radial transport forms an S-shape in the energy structure (L-E-plane) of the outer radiation belt. Low energy electrons (<300 keV) are less subject to hiss scattering below $L = 4$ Re. While, 0.3–1.5 MeV electrons are continuously depopulated between $L = 2.5 - 5$ Re.

Shprits et al. (2015) used the VERB-4D code to study electron behavior at transitional energies ($\mu \approx 100$ MeV/G) during the March 17th, 2013 storm using a constant outer boundary of 0.01–1.0 MeV at $L = 6.6$, based on Polar and CRRES satellite data. Their results show that convection alone cannot transport 100 MeV/G electrons to $L < 5$ Re regardless of geomagnetic activity levels, while convection and radial diffusion allow transport down to $L = 4$. Addition of local processes and losses due to chorus waves produces a radial increase in recovery phase fluxes. Shprits et al. (2015) present 3D simulations of this storm ignoring magnetopause losses. Since, the low energy and outer radial ($R = 6.6$ Re) boundaries of that study were inferred from GOES data, no electron tracing from the plasma sheet is involved in the estimation of the low energy population. Also, no sensitivity tests to the low energy boundary were presented, so that the role of the low energy electrons in the entire system cannot be examined. Furthermore, their simulation is strongly driven by the outer boundary, because $L = 6.6$ is very close to the heart of the outer belt.

In the present study, IMPTAM models the dynamics of plasma sheet electrons from the tail (outer radial boundary at $R = 9$ Re) to inner L-shell regions, free of satellite data. Using VERB-3D, we can account for losses to the magnetopause and study the coupling of different particle populations in the radiation belts, assuming a low energy population of $10 - 100$ keV, which is lower than in previous studies. For this study, we use statistical wave models in order to provide a more general validation of our coupled model that will help us extend it to a now-/forecasting tool. However, these wave models can produce results that might differ from previous studies on this particular storm.

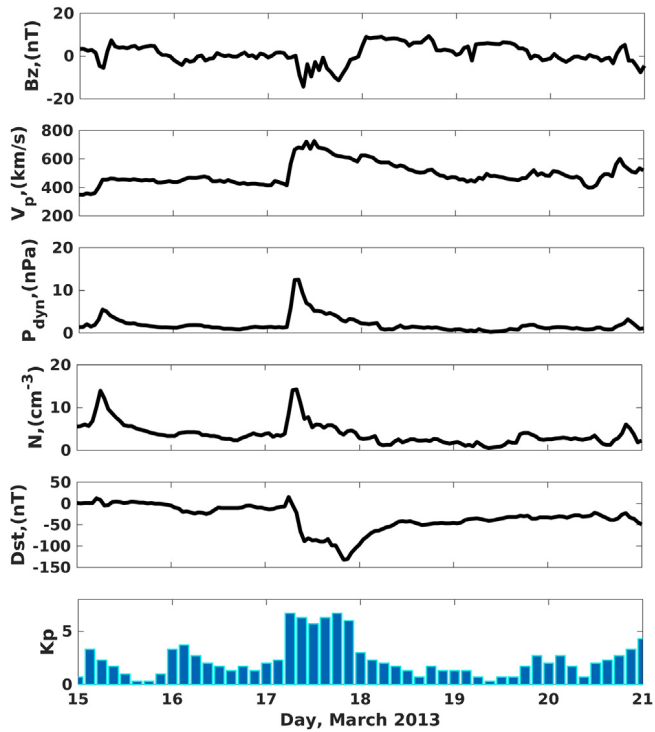

**Fig. 1.** Magnetic field and solar wind parameters, as well as geomagnetic indexes for the March 17th, 2013 storm. From top to bottom: vertical component of the IMF ($B_z$), solar wind velocity ($V_p$), dynamic pressure ($P_{dyn}$), number density (N), Dst and Kp indexes.
*Source:* Data from Omniweb.

Fig. 1 shows the vertical component of the magnetic field ($B_z$) (panel 1) and solar wind parameters (panels 2,3,4), together with the Dst and the Kp indexes (panels 5,6) for the time period March 15 − 20, 2013. A total of three geomagnetic events took place during this 6-day period. One during the first half of March 15th, the second at the beginning of March 16th and the third event (the studied event) starting early on the 17th of March. While the first two events had a duration of only a few hours, the main event lasted almost one entire day.

The event on March 15th has a rather minor magnitude as inferred from the north-south component of the IMF, which shows a moderate decrease becoming more southward by about (−5 nT). Increases in plasma velocity (to ∼450 km/s), dynamic pressure of the IMF (∼5.5 nPa) and proton density (∼14 cm⁻³) were measured. While the Kp-index clearly increases up to $Kp \approx 3^+$, Dst shows only minor variation during this event. The second event (on March 16th) can be better identified from the intensification of the Kp index to about $4^-$. Also, the Dst index diminishes to −25 nT. The variations of the remaining parameters are rather small. Although, of minor intensity, these storms might have produced electrons of keV energies that could increase fluxes at the heart of the outer belt.

Prior to the main geomagnetic storm, some time variations are observed in the solar wind parameters, IMF parameters and geomagnetic indexes. Parameters enhanced due to the first two minor storms, decreased and stabilized. On the day of the main storm (March 17th), a strong southward $B_z$-component of the IMF (∼ −14.4 nT) was measured. Sudden enhancements of the plasma velocity (up to about 720 km/s), plasma pressure (∼12.5 nPa) and proton density (∼14 cm⁻³) are observed. The north-south component of the IMF increases and oscillates for some hours, before decreasing in the afternoon to values below −11.5 nT. The Dst index shows two peaks of minima, Kp shows two peaks of maxima during this storm, similar to the $B_z$-component. Right at the beginning of the storm Dst drops to −89 nT and stays low for a couple of hours, before its values drop below −132 nT later that day. Simultaneously, the Kp-index reaches two maxima with values of $Kp \approx 7^-$.

### 3.1. Satellite data

In the current work, we use electron flux data for the studied period obtained from instruments on board both Van Allen Probes (Mauk et al., 2012) and the Geostationary Operational Environmental Satellites (GOES 13, 15) (Data Book GOES, 2005), covering the entire radial extent of the outer radiation belt. The two Van Allen Probes spacecrafts (initially named Radiation Belt Storm Probes (RBSP), RBSP A and B) have nearly identical orbits with perigee of ∼ 600 km altitude, apogee of 5.8 Re geocentric, and inclination of 10°, allowing their access to the most critical regions of the radiation belts, i.e. a radial range of $L = 2$ to $L \approx 5.8$ Re (Mauk et al., 2012). Each spacecraft hosts four identical Magnetic Electron Ion Spectrometers (MagEIS) (Blake et al., 2013): one low-energy unit (20 − 240 keV), two medium-energy units (80 − 1200 keV), and a high-energy unit (800 − 4800 keV). GOES 13 and 15 are American meteorologic satellites operated by the U.S. National Oceanic and Atmospheric Administration (NOAA) at nearly geosynchronous orbit. On board GOES 13 and 15, nine solid-state-detector telescopes (Magnetospheric Electron Detectors (MAGED)) provide pitch-angle resolved in-situ electron flux measurements over a radial range of $L = 5 − 6.6$ Re and in five energy bands: 30−50 keV, 50−100 keV, 100−200 keV, 200−350 keV and 350−600 keV. Five telescopes are oriented in the east–west (equatorial) plane and the other four telescopes in the north-south (meridional) plane (Hanser, 2011; Rodriguez, 2014a). Additionally, two Energetic Proton, Electron, and Alpha Detectors (EPEAD) on each satellite measure MeV electron and solar proton fluxes in two different energy ranges: >0.8 MeV and >2 MeV. One detector is oriented eastward and the other westward (Rodriguez, 2014b).

MAGED and EPEAD observations at a 5 min cadence are averaged over 30 min. EPEAD integral fluxes are obtained by averaging the measurements of the East and West telescopes, so that the resulting pitch-angles are also averages between both directions of the two telescopes (Rodriguez, 2014b). Integral fluxes as a function of energy are then fitted to a power law which is used to interpolate between values up to 1 MeV. For the conversion to differential flux, we use the 90° pitch-angle differential flux data from MAGED and fit the two integral channels of EPEAD (0.8 and 2 MeV) to an exponential function $f = A * exp(B * E)$, where f is the differential flux, E is the energy and A,B are time dependent coefficients calculated by solving the flux integral for averaged MAGED data. The pitch angle distribution below 500 keV is directly measured by MAGED. MagEIS pitch angle-resolved flux measurements from RBSP A and B are averaged over 30 min, and the energies are obtained from channel 5 (∼100 keV) of the low-energy unit together with channels 4 (∼400 keV) and 7 (∼900 keV) of the medium-energy unit. $L^*$ values were calculated using the magnetic field model (TS04) (Tsyganenko and Sitnov, 2005).

## 4. IMPTAM model results

### 4.1. Results of IMPTAM simulation

Fig. 2 presents the electron fluxes at geostationary orbit observed by the CEASE II ESA instrument onboard the AMC 12 satellite (thick black lines) and the electron fluxes modeled with IMPTAM for different energy ranges during the storm on March 15 − 20, 2013. The data are provided in the format of time-averaged differential fluxes (1/(cm² s sr eV)). The output of IMPTAM is the integral flux (1/(cm² s)) produced by all electrons coming from all directions with energies in the given energy ranges. In order to compare the observed and modeled fluxes accurately, we introduced the width of the energy channel and the solid angle $4\pi$ as follows: (modeled flux)/($4\pi\Delta E$).

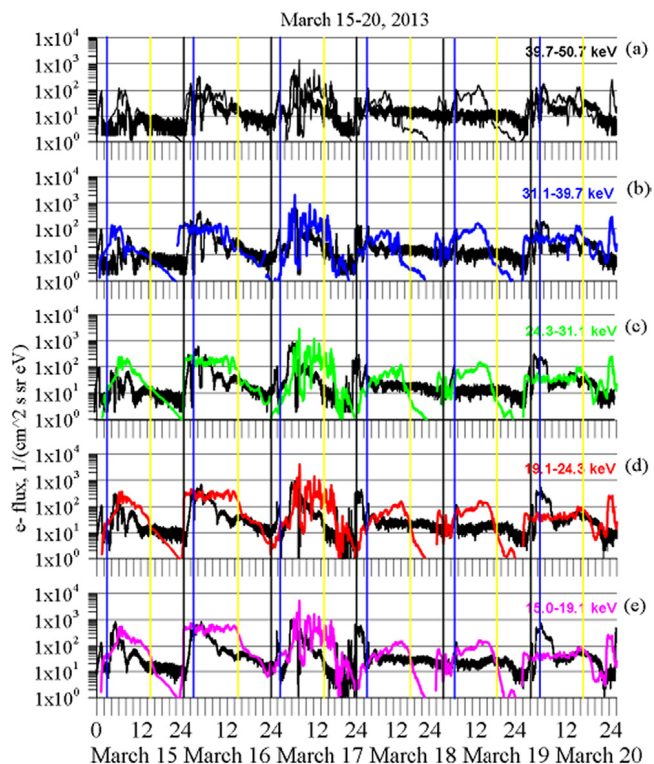

**Fig. 2.** Electron fluxes at geostationary orbit measured by the CEASE II ESA instrument onboard the AMC 12 satellite (thick black curves) and modeled with IMPTAM for: first panel 39.7 − 50.7 keV (black curve), second panel 31.1 − 39.7 keV (blue curve), third panel 24.3 − 31.1 keV (green curve), forth panel 19.1 − 24.3 keV (red curve), and fifth panel 15.0 − 19.1 keV (pink curve). The satellite's midnight (0230 UT) and noon (1430 UT) are marked with blue and yellow vertical lines, respectively. The end of each day is marked by black vertical lines.

For IMPTAM validation, we primarily use electron fluxes with energies from 15 to 50 keV and compare them to measurements of the instrument on board the geostationary satellite AMC12. This data set is used in order to conduct an independent validation of the IMPTAM performance, since the coupled IMPTAM-VERB model will be validated inside geostationary orbit using the data from Van Allen Probes and GOES.

AMC 12 geostationary satellite is at 322.5 degrees East and is equipped with the CEASE II (Compact Environmental Anomaly Sensor) instrument (Dichter et al., 1998), which contains an electrostatic analyzer (ESA). The CEASE II is a suite of various sensors intended to measure the in-situ space environment at the host spacecraft. The instrument contains a Lightly Shielded Dosimeter, a Heavily Shielded Dosimeter, a Particle Telescope (measuring high-energy electrons and protons), and an electrostatic analyzer for measuring low-energy electron fluxes (range 5 − 50 keV) in 10 channels.

As seen in Fig. 2, the modeled fluxes follow the general trends of the observations. During the main phase of the storm on March 17th, many variations are reproduced, but the modeled fluxes on the dayside are higher (the difference can reach an order of magnitude) than the observed ones. During the recovery phase of the storm on March 18th and 19th, the modeled fluxes drop much faster than the observed fluxes, when the satellite moves to the dusk sector via noon. This discrepancy may be due to inaccuracies in the parameterizations used for the electron losses.

Moreover, non-smooth transitions between different MLT-sectors and the simple combination of the electron lifetimes due to chorus and hiss waves can lead to deviations of the model predictions from the satellite data. Although the detailed dynamics of observed fluxes were not fully reproduced, this representation of electron lifetimes for keV electrons is currently the best available model. The keV electron fluxes vary significantly on the time scales of tens of minutes. Therefore, electron lifetimes parametrized by 3-hour Kp index do not reflect the full picture of shorter time variations.

Fig. 3 presents an example of the output of IMPTAM. It shows omnidirectional electron fluxes at 10 keV energy for March 17th, 2013, when the main phase of the storm occurred. Intense fluxes are observed at 08 UT, when Dst started dropping. Between $09 − 12$ UT and $15 − 20$ UT, the flux maps also exhibit clear increases at dawn. These intervals correspond to the times, when the two dips in Dst are observed. Afterwards, the electron fluxes start to decrease as the recovery phase progresses. This evolution is rather typical for the storm time variations of the keV electron fluxes.

### 4.2. Preparation of boundary conditions

#### 4.2.1. The low energy boundary

IMPTAM modeled fluxes were provided in a (L, MLT, $\alpha$, E) grid of ($31 \times 39 \times 23 \times 21$) points for each hour of the entire simulation period ($15 − 20$, March, 2013). The radial distance ($L$) has a range of 3 to 9 Re, the MLT (magnetic local time) range is 0° to 360°, equatorial pitch angles from 10° to 90° were computed and energies of 10 to 100 keV were modeled in a logarithmic scale. IMPTAM computed electron fluxes were averaged over all MLT sectors and fluxes at 90° pitch angle were extracted. For these particles the second adiabatic invariant ($J$) is zero, which will greatly simplify our calculations. Also in order to save computational time, a dipole approximation is used, i.e. we assume $L^* = L$. For future work, this should be corrected using $L^*$ tracing in a realistic magnetic field model as it might lead to discrepancies in the injection fluxes during storm time.

Further adaptation of the IMPTAM output to the VERB grid was done by extending modeled electron fluxes to radial distances below $L^* = 3$ at all energies, and by extending the energy range up to 1 MeV. As reported by Ripoll et al. (2017), flux magnitudes for energies $E > 100$ keV and $L^* < 3$ are rather low and decay very fast with increasing energy and decreasing $L^*$. Therefore, to extend IMPTAM fluxes, we have modeled a fast exponential decrease in $L^*$ and energy. The flux decrease in $L^*$ for $L^* < 3$ is given by:

$$\text{Flux}(L^* < 3, E) = \text{Flux}_{\text{IMPTAM}}(L^* = 3, E, 90°) * exp^{\frac{-(3-L^*)}{\Delta L}}) \quad (7)$$

the flux decrease in energy for $E > 100$ keV is given by:

$$\text{Flux}(L^* >= 3, E) = \text{Flux}_{\text{IMPTAM}}(L^*, E = 100 \text{ keV}, 90°)$$
$$* A * exp^{(\frac{\log_{10}(E) - \log_{10}(100 \text{ keV})}{\Delta E})} \quad (8)$$

where $\Delta L = 0.05$, $\Delta E = 0.19$ and $A = 0.22$ are the input coefficients. The main goal is to extract IMPTAM fluxes at $\mu_{min}$, which are the values relevant for the low energy boundary, and not to model the entire (L,E) plane. This approach might cut the radial extent of the outer belt below $L^* = 3$, which for energies below $\sim 300$ keV is not very precise. However, for energetic electrons at $\geq 400$ keV, this condition is valid. Finally, the processed IMPTAM output was interpolated to match the resolution of the VERB grid.

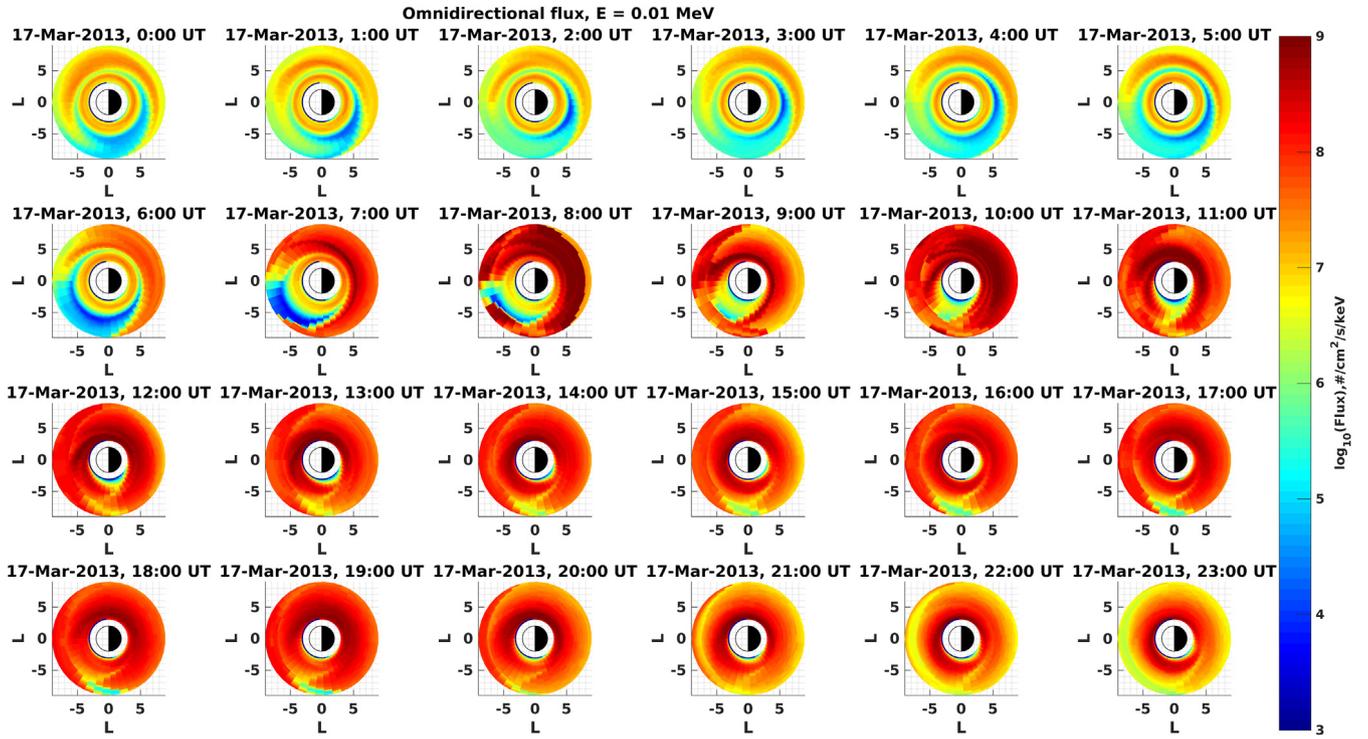

**Fig. 3.** Example of the IMPTAM output for the 17th of March, 2013. Displayed are omnidirectional electron fluxes at 10 keV energy for each hour of the day. Hour times are given in universal time (UT).

Fig. 4 presents an example of processed IMPTAM fluxes at 100 keV energy and 90° equatorial pitch angle (panel b), compared to satellite observations of GOES and Van Allen Probes (panel a) and the logarithmic difference (panel c) between panels a and b, calculated as $\log_{10}(\text{processed} - \text{IMPTAM}) - \log_{10}(\text{sat.data})$. Since, the fluxes displayed in this figure were treated for the coupling with VERB-3D, they do not represent a validation of IMPTAM.

The satellite observations (panel a) show that the lower $L^*$ edge of the outer belt is located between $L^* = 4 - 4.5$ and electron fluxes are rather low throughout March 15th and part of 16th. On the second half of March 16th, electron fluxes are moderately enhanced and the inner $L^*$ boundary of the outer belt moves below $L^* = 4$. The fluxes observed during the main storm undergo an abrupt increase of about two orders of magnitude and the lower $L^*$ boundary moves below $L^* = 2.5$. From the 17th to 19th, fluxes at all $L^*$ remain high with a peak around $L^* = 3$. On March 20th, initial flux decay becomes visible above $L^* = 3.5$. Processed IMPTAM fluxes (panel b) are in agreement with the observations prior to the storm onset and above $L^* = 5$ during the recovery phase. Pre-storm fluxes around $L^* = 3 - 5$ are overestimated by ~2 orders of magnitude. This is probably due to the used parametrizations of hiss-losses, which as reported by Ripoll et al. (2017) do not account for sufficient electron scattering during March 2013. In general, lifetimes are only a first order approximation and can yield important discrepancies to the realistic models. Also, storm-time injections and recovery phase fluxes around $L^* = 3.5$ show overestimation of about $1.5 - 2$ orders of magnitude. This might be partly due to used model of the source distribution in the plasma sheet, which is based on empirical relations with solar wind parameters (Dubyagin et al., 2016). Nevertheless, processed IMPTAM fluxes resemble the general electron dynamics observed in the satellite data.

The low energy boundary of the VERB code is located at $\mu_{min} = 9.3634$ MeV/G (see Section 2.2.2). We calculated $\mu$ for each point of the processed IMPTAM output at 90° pitch angle. Flux points matching the minimum $\mu$ condition ($\mu_{IMPTAM} = \mu_{min}$) are extracted together with their corresponding $L^*$ coordinate. Finally, in order to simplify the pitch-angle distribution for this initial study, we assumed a sinusoidal dependence of the pitch angles ($\sin^n(\alpha_0)$, for $n = 1$):

$$Flux_{BC}(L^*, \alpha_0) = Flux_{\text{IMPTAM}}(L^*, \alpha = 90°) \cdot \sin(\alpha_0). \quad (9)$$

where $\alpha_0$ is a pitch angle point in the VERB-grid. For the energies of the low energy boundary and particularly for $L > \sim 4.5$, this pitch-angle distribution might be too approximate (Shi et al., 2016). However, since pitch angle scattering occurs at very fast time scales and away from the boundary, the pitch-angle distribution at other grid points is defined by the shape of the $D_{\alpha\alpha}$ diffusion coefficient. Processed IMPTAM fluxes at the low energy boundary condition ($Flux_{BC}$) are then converted to PSD values ($PSD_{BC}$), as follows:

$$PSD_{BC} = Flux_{BC}(L^*, \alpha_0)/(pc)^2, \quad (10)$$

where $PSD_{BC}$ is the phase space density at the low energy boundary and only depends on $L^*$ and $\alpha_0$. These hourly PSD values are the input for the VERB-3D code.

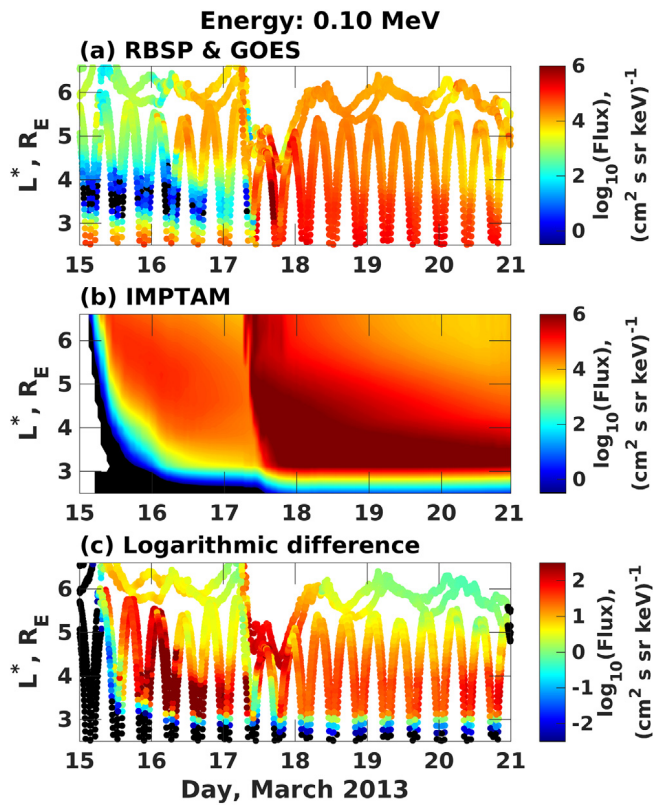

**Fig. 4.** Electron fluxes for 100 keV electrons at equatorial pitch angle $\alpha_0 = 90°$ as function of $L^*$ and time. (a) Satellite observations from MagEIS and MagED, (b) Electron fluxes computed with IMPTAM processed to match the computational grid of the VERB code, (c) logarithmic difference between IMPTAM simulated electron fluxes and satellite observations ($\log_{10}$(processed-IMPTAM)$-\log_{10}$(sat.data)).

### 4.2.2. The upper $L^*$ boundary for the VERB-3D code

The boundary condition at $L^* = 6.6$ provides the source of the low energy electrons (Subbotin et al., 2011a) in the VERB simulations. For the calculation of the upper $L^*$ boundary condition, we followed the approach of Brautigam and Albert (2000), who proposed scaling the electron flux distribution at the outer $L^*$ boundary with a boundary flux ($B_f$) to reproduce a realistic variation of the low energy electron fluxes.

The processing of the outer $L^*$ boundary is different from the processing of the low energy boundary. In this case, hourly MLT-averaged electron fluxes computed by IMPTAM at: $L^* = 6.6$, $E = 100$ keV, $\alpha_0 = 90°$ were extracted. Flux fluctuations at 100 keV were chosen, because this is the highest energy provided by IMPTAM. Using the long term electron flux spectrum ($Flux_{ave}$) measured by LANL satellites at $L^* = 7$, we can reconstruct the energy spectrum (up to 1 MeV) by scaling IMPTAM fluxes (Shprits et al., 2009b). For this, we calculate the coefficients of an exponential increase between the energy spectrum and the output of IMPTAM at three point energies (100 keV, 1 MeV and 3 MeV). Once these energies match the spectrum, electron fluxes between these reference energies are interpolated. With the estimated fluxes at 1 MeV, we calculate the scaling factor $B_f$, as follows:

$$B_f(t) = Flux_{\text{IMPTAM}}(t)/Flux_{ave}.$$

$B_f$ modulates the boundary fluxes of the VERB-3D simulations, as the initial PSD values calculated by the VERB code at $L^* = 6.6$ are multiplied by $(B_f(t))$ at each time step ($t$). The VERB-code then uses a statistical PSD-spectrum to scale PSD at $L^* = 6.6$ to other energies of the VERB-grid, thereby accounting for the seed population of a few 100s of keV. Those populations are transported radially according to the dynamics of the storm estimated by the Fokker–Planck equation (Eq. (5)). At this point, it is important to note that using 100 keV flux fluctuations might not fully represent the behavior of energetic boundary fluxes and that the scaling of PSD at the outer boundary with a statistical spectrum can lead to nonphysical increases of boundary fluxes. Also, the dropout of energetic electron fluxes (energy > 400 keV) occurring right after the injections (most obvious at 900 keV) is not observed at lower energies. Therefore, in our simulations, no dropouts will affect energetic fluxes prior to the main event.

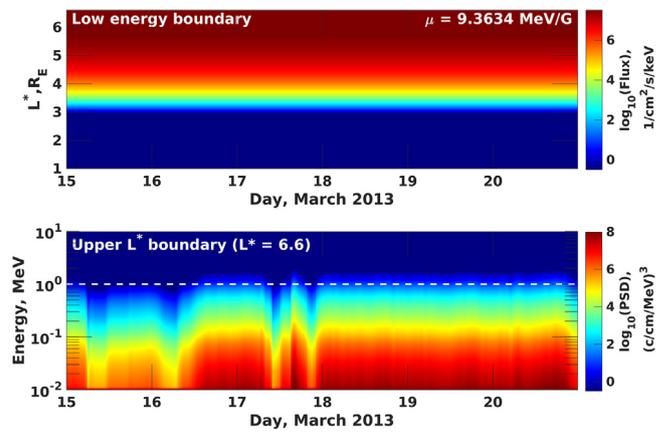

**Fig. 5.** Evolution of phase space density as function of time for the two boundary conditions used for a non-coupled VERB simulation. Upper panel: low energy boundary ($\mu = 9.3634$ MeV/G) calculated as the steady state solution of the radial diffusion equation. Lower panel: outer radial boundary ($L^* = 6.6$ at equatorial pitch angle $\alpha_0 = 85°$) estimated from satellite data after Drozdov et al. (2015). White dashed line marks 1 MeV energy.

## 5. Simulation results

Four different simulations are presented in this section:

I. A non-coupled VERB simulation using a constant low energy boundary condition and the outer $L^*$ boundary estimated from GOES data (Section 5.1).
II. A partially coupled IMPTAM-VERB simulation using the time dependent low energy boundary computed by IMPTAM and the outer $L^*$ boundary estimated from GOES data (Section 5.2).
III. A fully coupled IMPTAM-VERB simulation using both time dependent low energy and outer $L^*$ boundaries from IMPTAM without magnetopause losses (Section 5.3.1).
IV. A fully coupled IMPTAM-VERB simulation using the same boundary conditions as simulation III and accounting for losses to the magnetopause (Section 5.3.2).

The purpose of simulation I is to have a base for qualitative and quantitative comparison with the partially and fully coupled simulations. With simulation II, we test the sensitivity to the lower energy boundary. Simulation III shows the sensitivity of the fully coupled IMPTAM-VERB model to the upper $L^*$ boundary and simulation IV the effect of the losses to the magnetopause for this particular storm.

### 5.1. Non-coupled VERB simulation

A non-coupled VERB simulation was performed using the parameters described in Section 2.2. The low energy and outer $L^*$ boundary conditions are presented in Fig. 5 for particles at 85° pitch angle. Here, the low energy boundary condition (upper panel) was set constant in time and equal to the initial PSD value. At the upper radial boundary (lower panel), the initial PSD at $L^* = 6.6$ is multiplied by a time dependent scaling factor that accounts for GOES observations and the long-term PSD spectrum measured at $L^* = 7$.

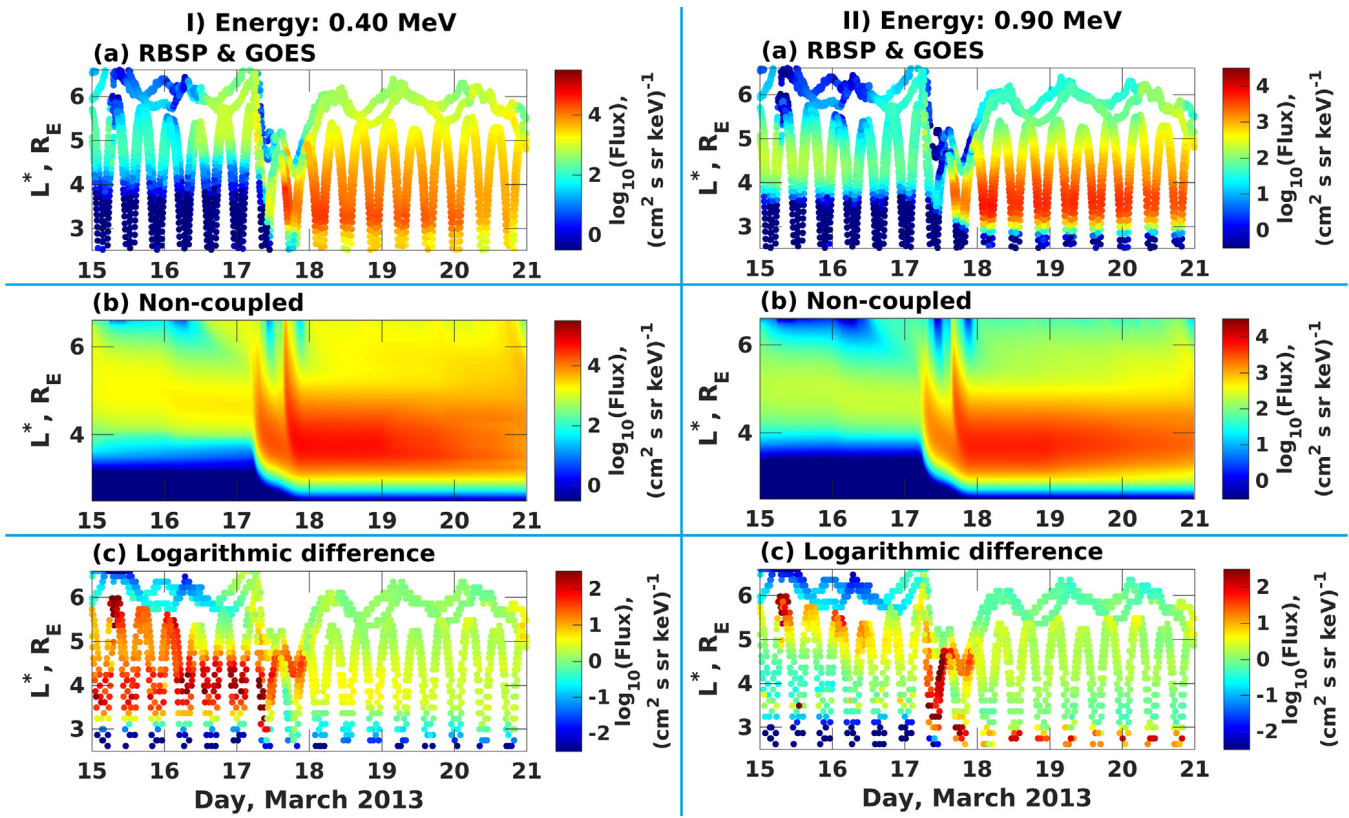

**Fig. 6.** Electron fluxes as a function of $L^*$ and time for electrons at equatorial pitch angle $\alpha_0 = 85°$ and fixed energies: (I) 0.40 MeV and (II) 0.90 MeV, respectively. The panels in each column show: (a) Van Allen probes and GOES data, (b) non-coupled VERB-3D simulation, (c) logarithmic difference between electron fluxes resulting from the non-coupled VERB-3D simulation and the satellite observations ($\log_{10}$(non-coupled)−$\log_{10}$(sat.data)). Note: color bars are different for each column.

On the first half of March 15th and 16th (Fig. 5, lower panel), the two minor storms produce small dropouts. Between these two small storms, electron PSD does not return to its initial levels. Prior to the main geomagnetic storm (March 17th), PSD at the boundary shows only minor variations. During the main phase of the storm, two intense particle injections and subsequent dropouts are observed at all energies. After the main phase, electron PSD shows a rather quite behavior.

Fig. 6 (panel a) shows satellite data from Van Allen probes and GOES, together with the resulting electron fluxes from the non-coupled VERB simulation (panel b) at different energies and the logarithmic difference (panel c) between panels a and b, calculated as $\log_{10}$(non-coupled)−$\log_{10}$(sat.data). At 400 keV, the simulations show overestimation of 0.5 to 2.5 orders of magnitude around $L^* = 3 - 6$ before the storm. Above $L^* = 6$ and below $L^* = 3$, underestimation of about 1 and 2.5 orders of magnitude is observed, respectively. Although, storm-time injection fluxes are higher by about 2 orders of magnitude between $L^* = 3 - 5$, fluxes are in agreement with the satellite data at higher $L^*$. During recovery phase, modeled electron fluxes match the data with 0.5 orders of magnitude accuracy. For 900 keV particles, the simulation results prior to the storm closely reproduce the satellite observations around $L^* = 3 - 5$. Overestimation of about one order of magnitude between $L^* \approx 5 - 6$ is observed, as well as underestimated fluxes at $L^* < 3$ and $L^* > 6$. Injection fluxes show values up to two orders of magnitude higher than the satellite data around $L^* = 3 - 5$. However, recovery phase fluxes are close to the observations at $L^* = 3 - 4$ and $L^* > 5$, showing overestimation of the radial extent of the belt at $L^* = 4 - 5$ (less than 1 order of magnitude) and at $L^* < 3$ (up to two orders of magnitude). Nevertheless, the non-coupled VERB simulation is able to reproduce the main features of the studied geomagnetic event, s.a. inward motion of fluxes during the main phase and flux buildup in the recovery phase.

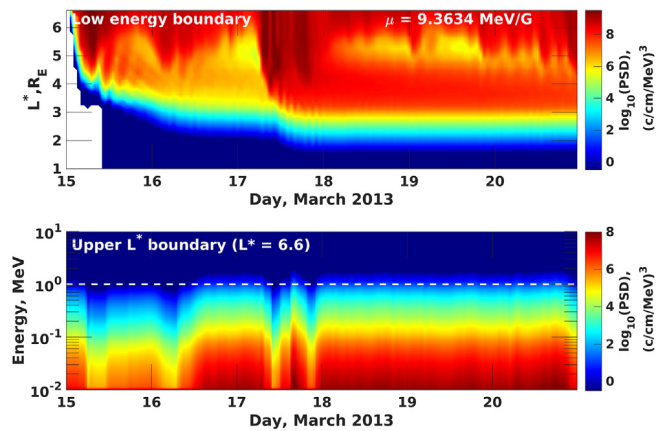

**Fig. 7.** Evolution of phase space density as a function of time for the two boundary conditions used for the partially coupled IMPTAM-VERB simulation. Upper panel: low energy boundary ($\mu = 9.3634$ MeV/G) provided from IMPTAM simulations. Lower panel: outer radial boundary ($L^* = 6.6$ at equatorial pitch angle $\alpha_0 = 85°$) estimated from satellite data after Drozdov et al. (2015). White dashed line marks 1 MeV energy.

### 5.2. Partially coupled simulation, low energy boundary from IMPTAM

Observations show that low energy fluxes can have a strong time variability (Jordanova and Miyoshi, 2005; Li et al., 2010). The input of fluxes simulated by IMPTAM at the low energy boundary of VERB simulations allows us to account for such variations and to study their effect on the outer radiation belt. In this partially coupled simulation, the low energy boundary was estimated by IMPTAM and the outer $L^*$ boundary computed from GOES observations (same as Fig. 5, lower panel). Evolution of PSD for both boundary conditions is presented in Fig. 7. The upper panel shows the low energy boundary extracted from IMPTAM simulations at $\mu \approx 9.4$ MeV/G, and the lower panel shows the upper $L^*$ boundary ($L^* = 6, 6$).

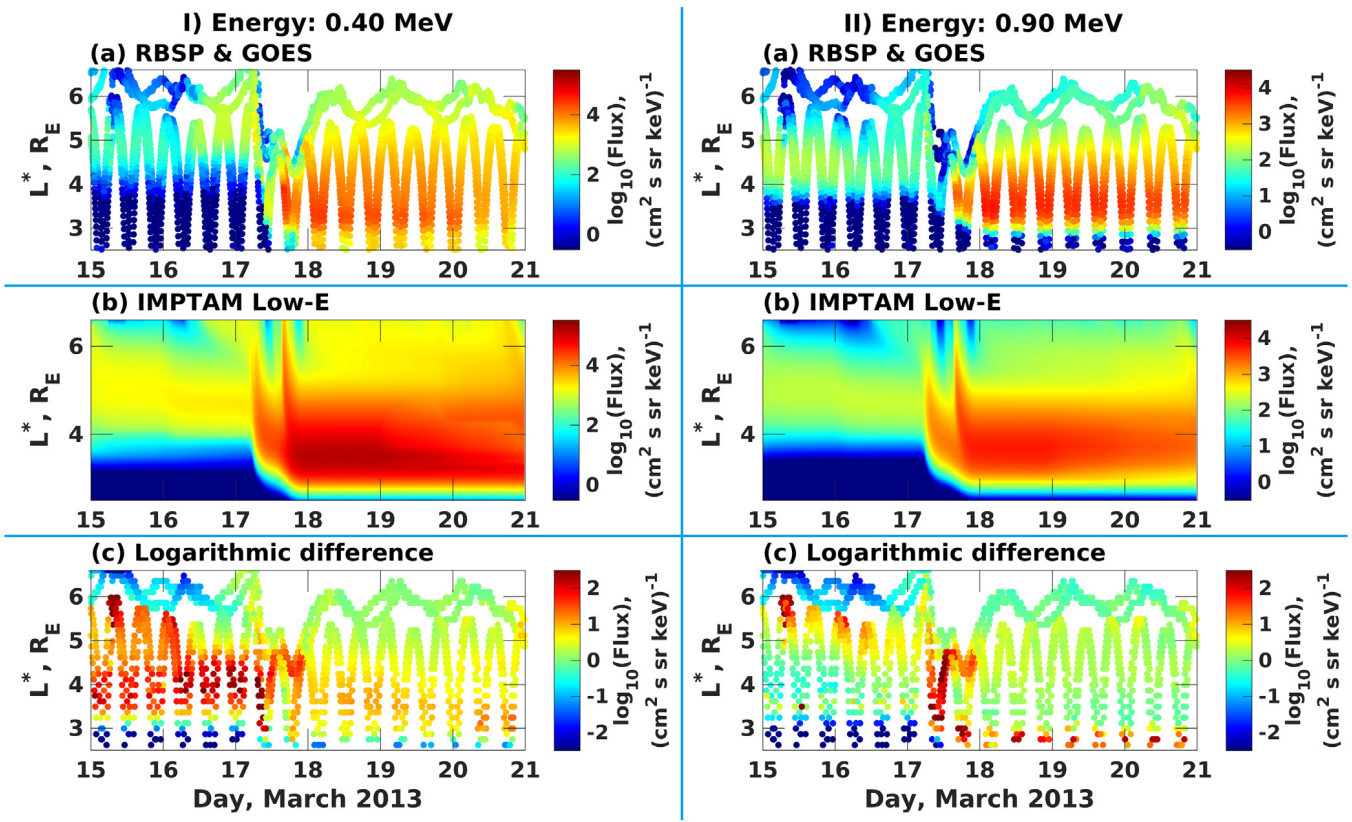

**Fig. 8.** Electron fluxes as a function of $L^*$ and time for equatorial pitch angle $\alpha_0 = 85°$ at fixed energies: (I) 0.40 MeV and (II) 0.90 MeV, respectively. The panels in each column show: (a) Van Allen probes and GOES data, (b) partially coupled IMPTAM-VERB simulation using the low energy boundary provided by IMPTAM and the upper $L^*$ boundary from GOES data, (c) logarithmic difference between electron fluxes resulting from the partially coupled IMPTAM-VERB simulation and the satellite observations ($\log_{10}(\text{part.coupled}) - \log_{10}(\text{sat.data})$). Note: color bars are different for each column.

IMPTAM simulations start with an empty magnetosphere that is continuously filled up and stabilized during the initial hours of the simulation. For this reason, the PSD observed at the beginning of the simulation has very low values. After a moderate enhancement during the first half of March 15th, PSD values start increasing gradually and reaching lower L-shells, $L^* \approx 4$. On March 15th and 16th, minor injections are also observed. These two events are well correlated with the depletions observed in the upper $L^*$ boundary from satellite data and generate minor peaks in PSD around $L^* \approx 4$, indicating particle transport to inner regions that will further affect the VERB simulations. During the main storm, a series of particle injections from higher to lower L-shells is simulated by IMPTAM over several hours. Two moderate injections are observed before 12:00 UT and a more intense particle injection is modeled around $18:00 - 20:00$ UT, which is consistent with the findings of Yu et al. (2014). During these time intervals, radial transport is enhanced and the PSD increases by more than two orders of magnitude around $L^* = 3$. During the recovery phase PSD values close to geosynchronous orbit return to quite time levels, but the peak at $L^* = 3$ remains throughout the rest of the simulation decaying slowly.

Fig. 8 displays satellite data (panel a) for different energies (different columns), the corresponding results of the partially coupled simulation (panel b) and the logarithmic difference (panel c) between panels a and b, calculated as $\log_{10}(\text{part.coupled}) - \log_{10}(\text{sat.data})$. Simulations of 400 keV particles resemble the general flux evolution of the observations, but show overestimation of one to two orders of magnitude between $L^* = 3 - 6$ during March 15th. One day later, fluxes above $L^* = 5$ match the observations and between $L^* = 3 - 5$ show overestimation of about 2 orders of magnitude. Prior to the storm, flux values at $L^* < 3$ and $L^* > 6$ are underestimated by 1 to 2 orders of magnitude. During the main phase, overestimation of about 1 order of magnitude is observed locally between $L^* = 2.5 - 5$. During the recovery phase, fluxes above $L^* = 5$ are close to the satellite data, but overestimation of about 1 order of magnitude is modeled at the heart of the belt. On the other hand, fluxes at 900 keV have a very similar evolution to the non-coupled 900 keV electron fluxes. In the partially coupled simulation, only slightly higher flux values (less than 0.5 orders of magnitude) are observed between $L^* = 3 - 4$ on March 18th. During the recovery phase, minor increase of fluxes occurs below $L^* = 3$ (also of less than 0.5 orders of magnitude), probably due to insufficiency of the modeled flux decrease (at $L^* < 3$) to represent realistic fluxes in the slot region. Since the only difference between the non-coupled and partially coupled simulations is the low energy boundary condition from IMPTAM, this sensitivity test indicates that the dynamics of MeV electrons have only a minor or negligible dependence on the electron population of $10 - 100$ keV energies (i.e. $\mu = 9.3634$ MeV/G), which is consistent with the conclusions of Subbotin et al. (2011a). Furthermore, 400 keV particle dynamics close to the heart of the belt appear to be linked to the evolution of IMPTAMs low energy boundary, i.e. about $50 - 100$ keV between $L^* = 3 - 5$. These electron populations provide enough energy to accelerate 400 keV electrons, but not 900 keV electrons.

### 5.3. Fully coupled simulation, low energy and outer $L^*$ boundaries from IMPTAM

Finally, to complete the coupling of IMPTAM and VERB, the source of the low energy electron population at geostationary orbit was accounted for, by introducing IMPTAM-simulated electron PSD at the outer $L^*$ boundary ($L^* = 6.6$) of the VERB simulations. The electron PSD used at the low energy boundary is the same as in the partially coupled simulation (Section 5.2).

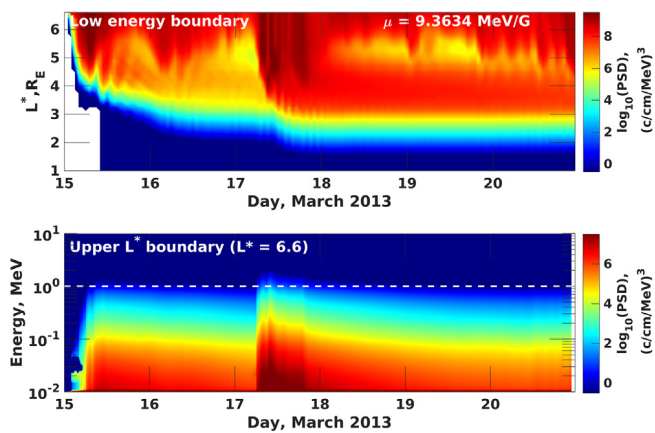

**Fig. 9.** Evolution of electron PSD as a function of time for the two boundary conditions used in the fully coupled IMPTAM-VERB simulation. Upper panel: low energy boundary ($\mu = 9.3634$ MeV/G) provided from IMPTAM simulations. Lower panel: outer radial boundary ($L^* = 6.6$ at equatorial pitch angle $\alpha_0 = 85°$) estimated using IMPTAM simulated fluxes. White dashed line marks 1 MeV energy.

Fig. 9 presents the time varying PSD used at the two boundaries in question, the upper panel shows the low energy boundary (same as in Fig. 7, upper panel) and the lower panel displays the upper $L^*$ boundary. At the outer $L^*$ boundary, moderate injections of low energy particles (max. 100 keV) are observed during the first half of March 15th. These were triggered by the small geomagnetic event taking place earlier that day and although transport of high energy particles is also involved here, the PSD for energies above 100 keV is low to moderate, and around 1 MeV even very low. After this initial storm, PSD values decrease at all energies. During the main phase, two subsequent increases in PSD occur, indicating motion of low and high energy particles from the upper $L^*$ boundary to lower L-shells. Electron PSD for up to some 100s of keV energies is high to moderate, while for particles with higher energy, the PSD values are rather low. During the recovery phase, the PSD maintains higher levels than those of quiet times for several hours, before it starts decaying on March 18th.

### 5.3.1. Fully coupled simulation, sensitivity to the outer $L^*$ boundary

Electron fluxes resulting from the fully coupled IMPTAM-VERB simulation are presented in Fig. 10 (panel b) for different energies (different columns), together with the corresponding satellite observations of GOES and Van Allen Probes (panel a) and the logarithmic difference (panel c) between panels a and b, calculated as $\log_{10}$(full-coupled)$-\log_{10}$(sat.data). The coupled model is able to reproduce well the general evolution of electron fluxes throughout the simulation. At 400 keV energies, storm-time fluxes and the maximum peak around $L^* = 3.5$ during the recovery phase show overestimation of about one order of magnitude. Enhanced radial transport leads to an overestimation of recovery phase fluxes above $L^* = 4$ of ∼0.5 − 1 orders of magnitude, generating a broader radial extent of the belt. Electron fluxes at 900 keV are in the same orders of magnitude as the measurements at $L^* = 3 − 4.5$ prior to the storm and are within 0.5 orders of magnitude at $L^* = 3 − 4$ during the recovery phase. Overestimation of the radial extent of the belt (about 1 order of magnitude) is observed between $L^* = 4 − 5$.

Comparison with the non-coupled and partially coupled simulations shows the strong influence of the outer $L^*$ boundary condition on the VERB simulations. Fluxes resulting from the fully coupled simulation show higher fluxes than those of the non-coupled simulation. Compared to the partially coupled simulation, an increase of the radial extent of the belt is observed during the recovery phase enhancing fluxes around $L^* = 4 − 5$ and the peak around $L^* = 3.5$ by about 1 order of magnitude. Fluxes above $L^* = 6$ decrease by about 0.5 orders of magnitude with respect to the partially coupled simulations.

Although, the outer $L^*$ boundary from IMPTAM and the one from GOES data are in the same orders of magnitude, their time evolution is not quite the same. Also, due to the complexity of the models, there could be several reasons for the observed differences between the satellite data and the three presented simulations: (1) The injections simulated on March 15th and 16th are overestimated by IMPTAM. This might be related to limitations in the source distribution model of the plasma sheet incorporated in IMPTAM. (2) The processing performed on IMPTAM fluxes, necessary for the code coupling, could have introduced biases in the boundary conditions. (3) Lifetime parametrizations included in IMPTAM may not be realistic enough to account for the entire electron losses. (4) Also, losses to the magnetosphere were not included in the simulations.

Since the outer $L^*$ boundary strongly influences VERB simulations, the dynamics simulated by IMPTAM should be as accurate as possible. The upper $L^*$ boundaries from IMPTAM and from GOES measurements at $L^* = 6.6$ are in the same orders of magnitude. Injections on the first two days of the simulation are present in both boundaries, but in IMPTAM show some overestimation that generated a considerable amount of low energy electrons inside the outer belt. This is partly because the model of the source distribution in the plasma sheet embedded in IMPTAM is based on empirical relations with solar wind parameters (Dubyagin et al., 2016), which could lead to the observed inaccuracies of injection fluxes. Overestimation is also partly caused by the initial condition of the VERB-3D simulations, which was chosen to match the satellite observations at 900 keV energy. However, since the overestimation of fluxes is also observed in the non-coupled simulations (at 400 keV) and in Fig. 4, this could be an indication that the hiss wave model is too weak at low energies and therefore not able to wipe out these low energy electrons around $L^* ≈ 4$. While in the model of Orlova et al. (2014), diffusion rates applied for energies below 500 keV at $L^* = 3 − 4$ are on the order of 1 − 10 days, Ripoll et al. (2017) show that prior to the main storm (March 15th and 16th) loss rates are on the order of ∼3 hours around $L^* = 2.5 − 4$ Re.

Another possible source of overestimation could be the use of parametrized losses in IMPTAM. Lifetime parametrizations are often used to account for particle losses due to wave-particle interactions, because of their rather simple implementation in physics-based codes. However, lifetimes are only a first order approximation and can yield important discrepancies with the realistic models. Ripoll et al. (2017) reported that the combination of radial diffusion coefficients from Brautigam and Albert (2000) and the parametrized losses from Orlova et al. (2014) tend to overestimate the radial extent of the outer belt at all energies. However, Ripoll et al. (2017) did not account for losses to the magnetopause, which are important for this storm. Also, the fluctuations of the outer $L^*$ boundary are extracted from the dynamics of 100 keV electrons, as this was the highest energy provided in the IMPTAM grid. Such an approach assumes a similar behavior of low and high energy populations and can, therefore, lead to inaccuracies. Extending the computational grid of IMPTAM to energies of a few 100s of keV would allow a more realistic estimation of the outer $L^*$ boundary condition. From a numerical perspective, the processing applied to the output of IMPTAM included modeling an exponential decrease of high energy electron fluxes below $L^* = 3$. This approach might be too approximate to describe the dynamics of such particles. Also, linear interpolation between grids can add biased values to the boundary conditions.

### 5.3.2. Fully coupled simulation, losses to the magnetopause

Higher fluxes in our simulations at all energies indicate the lack of effective loss mechanisms that could balance enhanced inward transport. Hudson et al. (2015) reported in detail the importance of magnetopause shadowing during this storm, which through strong outward radial transport probably generated the dropout observed during storm onset. In order to address the effect of magnetopause compression in our simulations, we have modeled the magnetopause location as the last closed drift shell (LCDS) on the day side for each day of the simulation. Using the International Radiation Belt Environment Modeling (IRBEM) library (Boscher et al., 2013), we estimated pitch-angle dependent LCDSs in the TS04 magnetic field model (Tsyganenko and Sitnov, 2005) (Fig. 11). To model the losses, we assume that all particles on a certain $L^*$ are lost once the magnetopause crosses this L-shell, i.e. for all $L^* \geq$ LCDS, PSD is set equal to zero. This approximation can only be regarded as a worse case scenario, because in general not all electrons are transported out of the magnetosphere during magnetopause compression.

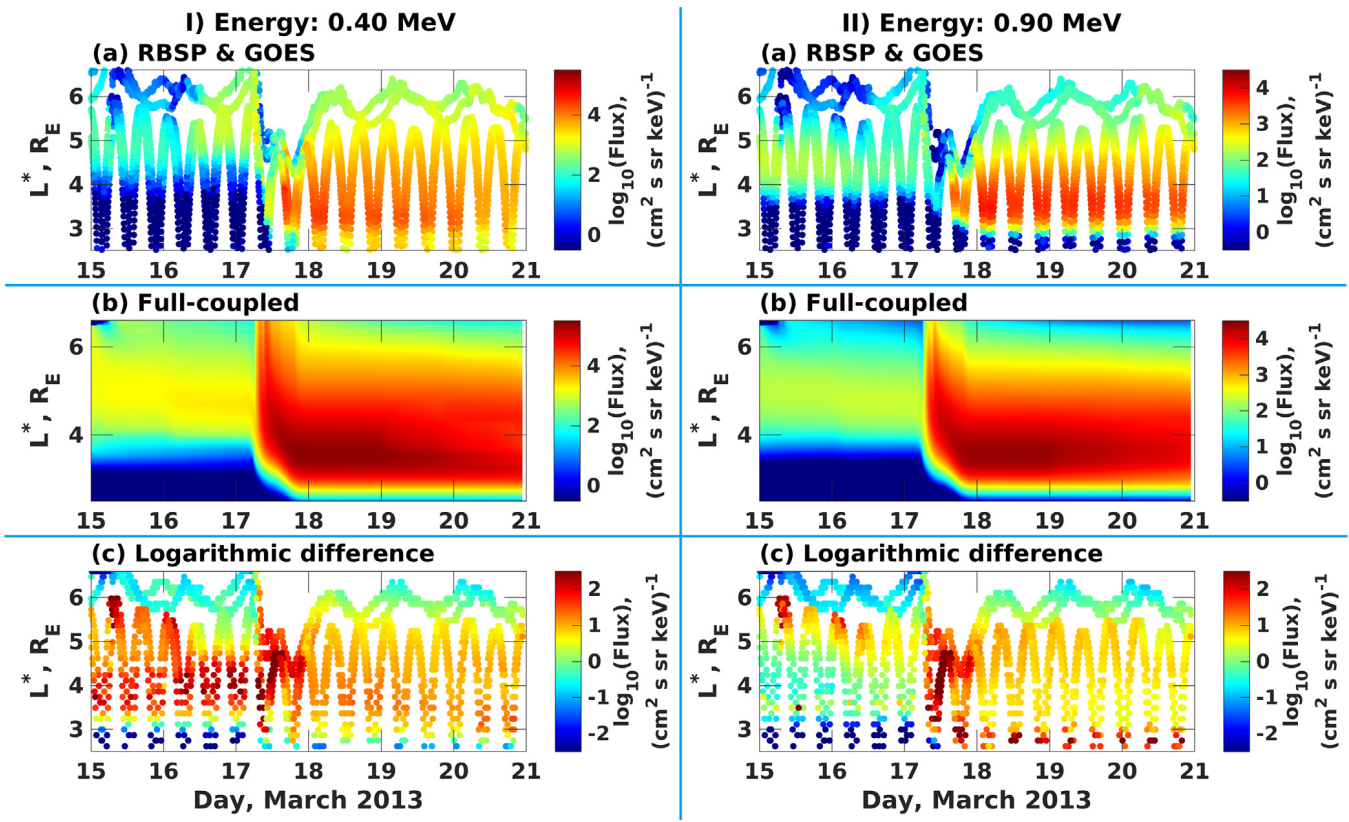

**Fig. 10.** Electron fluxes as a function of $L^*$ and time for equatorial pitch angle $\alpha_0 = 85°$ at fixed energies: (I) 0.40 MeV and (II) 0.90 MeV, respectively. The panels in each column show: (a) Van Allen probes and GOES data, (b) fully coupled IMPTAM-VERB simulation using the low energy and outer $L^*$ boundary provided from IMPTAM, (c) logarithmic difference between electron fluxes resulting from the fully coupled IMPTAM-VERB simulation and the satellite observations ($\log_{10}$(full-coupled)− $\log_{10}$(sat.data)). Note: color bars are different for each column.

Fig. 12 presents the results (panel b) of a coupled IMPTAM-VERB simulation accounting for magnetopause losses, as previously described, and using the boundary conditions presented in Fig. 9. Similar to previous figures, satellite data (panel a) and the logarithmic difference (panel c) between the simulation and the satellite observation, calculated as $\log_{10}$(full-coupled + MP)− $\log_{10}$(sat.data), are also presented. Direct comparison of these results with the full coupled simulation in Fig. 10 enlightens the role of enhanced outward radial diffusion for this specific event. Fluxes of 400 keV energies do not present major changes during pre-storm times, apart from a few localized minor reductions at higher L-shells. However, after storm onset fluxes between $L^* = 3.5 - 5.5$ drop by about 1 to 2 orders of magnitude and above $L^* = 5.5$ by more than two orders of magnitude. Although, recovery phase fluxes at $L^* > 4$ agree with the satellite observations, between $L^* = 3 - 4$ overestimation of less than one order of magnitude is observed. Flux evolution of 900 keV particles changes drastically with the implementation of magnetopause losses. At $L^* > 6$, some losses are observed on the days prior to the storm. These losses have only a limited effect on fluxes at lower L-shells ($L^* = 5 - 6$), but a reduction of about 1 order of magnitude is reproduced on the 15th. During main and recovery phases a decrease in fluxes of up to 1.5 orders of magnitude occurs at $L^* \geq 3$, compared to the full-coupled simulation.

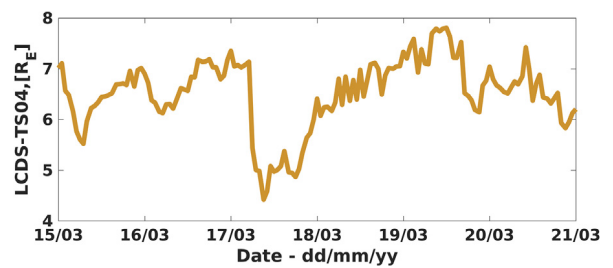

**Fig. 11.** Magnetopause location for electrons with equatorial pitch angle $\alpha_0 = 85°$, calculated as the last closed drift shell (LCDS) using Tsyganenko model (TS04) (Tsyganenko and Sitnov, 2005).

Similar to the small particle injections on the 15th and 16th, main storm injections appear to be overestimated by about one order of magnitude. This is probably due to the scaling approach used for boundary fluxes and the fact that the dropout of energetic electron fluxes (energy > 400 keV) occurring right after the injections (most obvious at 900 keV) is not observed at lower energies. However, the timing of IMPTAMs boundary injections is in agreement with the satellite data and reproduces well the dynamics at lower $L^*$. Our simulations also show time delays for flux enhancement at $L^* \approx 4$, consistent with those reported by Boyd et al. (2014). In general, accounting for losses to the magnetopause leads to an agreement of the simulated belt and the satellite measurements within one order of magnitude at 900 keV during storm and recovery phases and for recovery fluxes at 400 keV between $L^* = 3 - 6$. Above $L^* = 6$, resulting fluxes show some underestimation, indicating that the assumed magnetopause losses are stronger than the real loss mechanism. Interestingly, the full-coupled simulation with magnetopause losses and the non-coupled simulation, which used satellite data, achieve modeling results within 0.5 orders of magnitude accuracy during main- and recovery phases.

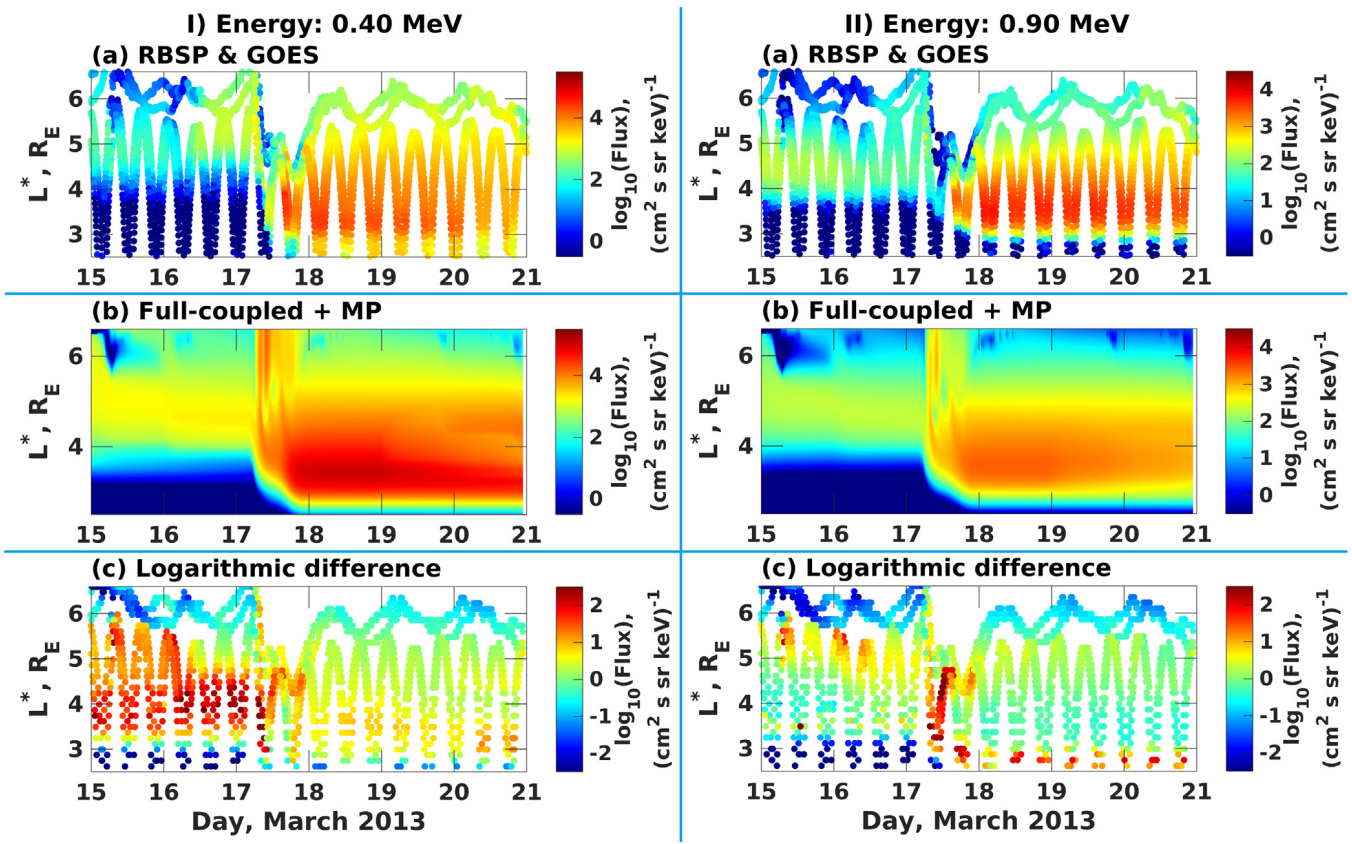

**Fig. 12.** Electron fluxes as a function of $L^*$ and time for equatorial pitch angle $\alpha_0 = 85°$ at fixed energies: (I) 0.40 MeV and (II) 0.90 MeV, respectively. The panels in each column show: (a) Van Allen probes and GOES data, (b) fully coupled IMPTAM-VERB simulation using the low energy and outer $L^*$ boundary provided from IMPTAM and including magnetopause losses, (c) logarithmic difference between electron fluxes resulting from the fully coupled IMPTAM-VERB simulation (with magnetopause losses) and the satellite observations ($\log_{10}$(full-coupled + MP)$-\log_{10}$(sat.data)). Note: color bars are different for each column.

## 6. Summary and conclusions

Low energy electrons (10s to a few 100s of keV) in the inner magnetosphere can be accelerated by combined convection, radial transport and local acceleration, thus influencing electron dynamics in the radiation belts. For this reason, it is important to account for realistic dynamics of the low energy electron population to generate accurate time dependent models of the radiation belt region. In this study, we couple IMPTAM and VERB-3D to obtain a radiation belt model that accounts for realistic dynamics of the low energy electron population. The behavior of the low energy population is calculated using the IMPTAM code, which traces electrons (10–100 keV) from the plasma sheet to inner L-shells and computes electron dynamics due to magnetospheric convection and radial diffusion. Losses due to pitch angle scattering caused by wave-particle interactions with chorus and hiss waves are included in the model. The output of IMPTAM is used at the low energy and outer $L^*$ boundary conditions of the VERB-3D code simulations. The VERB code calculates the evolution of energetic electrons in the radiation belts accounting for diffusion processes and a number of wave particle interactions. The IMPTAM-VERB coupled model does not require satellite data to perform simulations and is therefore suitable for forecasting purposes.

The IMPTAM-VERB coupled model was tested on a six-day simulation period from March 15th to 20th, 2013, in order to include the main development and phases of the March 17th, 2013 storm. Electron fluxes computed by IMPTAM are overestimated in the days prior to the storm, probably due to inaccuracies in the used loss parametrizations and in the plasma sheet source distribution model. Flux injections during the main event and the subsequent peak around $L^* = 3 - 4$ are well reproduced, although with some overestimation. In general, IMPTAM simulations are in reasonable agreement with the GOES and RBSP data. The output of IMPTAM was processed to match the computational grid of the VERB code and used as the boundary conditions of VERB simulations. **Four different simulations** were performed: **(1)** A non-coupled VERB simulation, using a constant low energy boundary condition and the outer $L^*$ boundary estimated from GOES data; **(2)** A partially coupled IMPTAM-VERB simulation, using the time dependent low energy boundary computed by IMPTAM and the outer $L^*$ boundary estimated from GOES data; **(3)** A fully coupled IMPTAM-VERB simulation, using both time dependent low energy and outer $L^*$ boundaries computed by IMPTAM; **(4)** A fully coupled IMPTAM-VERB simulation, using the same set up as in simulation 3 and including losses to the magnetopause. A summary of the simulation results is given below:

- **Simulation (1):** The purpose of the non-coupled VERB simulation was to have a basis for comparison with the partially and fully coupled simulations. Electron fluxes at 400 keV show overestimation of about 2 orders of magnitude during storm time ($L^* = 3-5$), but are agree with the data within 0.5 orders of magnitude during the recovery phase above $L^* = 3$. At 900 keV, fluxes are in agreement with the satellite data within 0.5 orders of magnitude for most part of the simulated period.

- **Simulation (2):** The partially coupled simulation was performed to test the sensitivity of the coupled model to the low energy boundary. At 400 keV, flux overestimation of less than 1 order of magnitude is seen in main and recovery phase fluxes around $L^* = 4$. However, despite the differences between the low energy boundaries of simulations 1 and 2, electron fluxes at 900 keV for these two simulations are very similar and agree with the observations within 0.5 orders of magnitude during pre-storm times and recovery phase above $L^* = 3$.
- **Simulation (3):** This simulation represents the fully coupled IMPTAM-VERB model and its sensitivity to the upper $L^*$ boundary. Although, flux injections during the main phase are overestimated causing a broader radial extent of the belt than in previous simulations, the model reproduces well the general trends and enhancements of electron fluxes throughout the simulation period. At 400 and 900 keV, modeled fluxes show similar magnitudes to the fluxes of simulation 1 and 2 during pre-storm times, and an increase in main and recovery phase fluxes of about 1 order of magnitude is observed below $L^* = 6$.
- **Simulation (4):** This simulation includes magnetopause losses in the fully coupled IMPTAM-VERB model. At 400 keV, fluxes below $L^* = 6$ are strongly reduced compared to simulation 3 and also the radial extent of the belt. Fluxes around $L^* = 4$ are overestimated by less than 1 order of magnitude. Losses to the magnetopause reduces fluxes by up to 2 orders of magnitude in comparison to simulation 3 (for 900 keV) and decreases the radial extent of the belt by about 1 Re. Modeling results of simulations 1 and 4 have similar levels of accuracy (within 0.5 orders of magnitude compared to sat. data), although simulation 4 does not use satellite measurements.

For this particular storm, the simulations with the IMPTAM-VERB coupled model lead to three main conclusions: (1) while the low energy population (max. energy 100 keV) seems to affect the dynamics of electrons up to 400 keV energies, high energy particle (>900 keV) dynamics appear to be rather insensitive to this population; (2) unlike the low energy boundary, the outer $L^*$ boundary does have a significant influence on energetic electron dynamics; (3) magnetopause losses reduce storm time influx by up to two orders of magnitude. Although, the approach used to estimate the magnetopause losses underlies a strong assumption, the model is able to reproduce the flux evolution at 900 keV with high accuracy (within 0.5 orders of magnitude) below $L^* = 6$. Moreover, this study demonstrates that the IMPTAM-VERB coupled model provides a reliable satellite-data-independent tool for the modeling of radiation belt dynamics.

The IMPTAM-VERB coupled model could be improved in a number of ways. Increasing the accuracy of the plasma sheet source distribution model could affect the simulations considerably. Extending the energy range of IMPTAM would also increase accuracy of the outer $L^*$ boundary condition. The coupled IMPTAM-VERB model is completely independent of satellite data and only depends on solar wind parameters, Dst and Kp indexes. Therefore, it can be easily used as a forecasting tool when the driving parameters are known in advance, e.g. from the NOAA solar wind predictions of the Wang–Sheeley–Arge(WSA)-ENLIL model (Parsons et al., 2011), from the Space Weather Modeling Framework (SWMF) (e.g. Haiducek et al., 2017; Liemohn et al., 2018; Wintoft and Wik, 2018) or from the SWIFT-code (Arber et al., 2001).

## Acknowledgments

The project leading to these results has received funding from the European Union's Horizon 2020 research and innovation program under grant agreement No. 637302 PROGRESS, Germany and No. 776287 SWAMI. The work of A. Castillo has been funded by the Deutsche Forschungsgemeinschaft (DFG), Germany under grant agreement CRC 1294, Project B06. The work of N. Ganushkina at the University of Michigan was partly supported by the National Aeronautics and Space Administration, United States under grant agreement NNX17AI48G and by the National Science Foundation, United States under grant agreement NSF 1663770. The work of S. Dubyagin and, partly, of N. Ganushkina leading to these results has been carried out in the Finnish Centre of Excellence in Research and Sustainable Space (Academy of Finland grant numbers 312351 and 312390), which we gratefully acknowledge. The authors used geomagnetic indices provided by OMNI-Web (http://omniweb.gsfc.nasa.gov/form/dx1.html) and are grateful to the RBSP-ECT team for the provision of Van Allen Probes observations (http://www.rbsp-ect.lanl.gov/). We sincerely acknowledge Adam Kellerman for the preparation of the used satellite data. In addition, the authors are appreciative of the valuable comments from the reviewers.

## References

Abel, B., Thorne, R.M., 1998. Electron scattering loss in earth's inner magnetosphere: 1. Dominant physical processes. J. Geophys. Res. Space Phys. 103 (A2), 2385–2396.

Agapitov, O., Artemyev, A., Krasnoselskikh, V., Khotyaintsev, Y.V., Mourenas, D., Breuillard, H., Balikhin, M., Rolland, G., 2013. Statistics of whistler mode waves in the outer radiation belt: Cluster staff-sa measurements. J. Geophys. Res. Space Phys. 118 (6), 3407–3420.

Albert, J., 2007. Simple approximations of quasi-linear diffusion coefficients. J. Geophys. Res. Space Phys. 112 (A12).

Albert, J., 2013. Comment on "on the numerical simulation of particle dynamics in the radiation belt. Part I: Implicit and semi-implicit schemes" and "on the numerical simulation of particle dynamics in the radiation belt. Part II: Procedure based on the diagonalization of the diffusion tensor" by E. Camporeale et al. J. Geophys. Res. Space Phys. 118 (12), 7762–7764.

Albert, J.M., Meredith, N.P., Horne, R.B., 2009. Three-dimensional diffusion simulation of outer radiation belt electrons during the 9 october 1990 magnetic storm. J. Geophys. Res. Space Phys. 114 (A9).

Albert, J., Young, S., 2005. Multidimensional quasi-linear diffusion of radiation belt electrons. Geophys. Res. Lett. 32 (14).

Ali, A.F., Malaspina, D.M., Elkington, S.R., Jaynes, A.N., Chan, A.A., Wygant, J., Kletzing, C.A., 2016. Electric and magnetic radial diffusion coefficients using the Van Allen probes data. J. Geophys. Res. Space Phys. 121 (10), 9586–9607.

Arber, T., Longbottom, A., Gerrard, C., Milne, A., 2001. A staggered grid, Lagrangian-Eulerian remap code for 3-d MHD simulations. J. Comput. Phys. 171 (1), 151–181.

Baker, D., Allen, J., Belian, R., Blake, J., Kanekal, S., Klecker, B., Lepping, R., Li, X., Mewaldt, R., Ogilvie, K., et al., 1996. An assessment of space environmental conditions during the recent anik e1 spacecraft operational failure. ISTP Newslett. 6 (2), 8.

Baker, D., Stone, E., 1978. The magnetopause energetic electron layer, 1. Observations along the distant magnetotail. J. Geophys. Res. Space Phys. 83 (A9), 4327–4338.

Birn, J., Thomsen, M., Borovsky, J., Reeves, G., McComas, D., Belian, R., 1997. Characteristic plasma properties during dispersionless substorm injections at geosynchronous orbit. J. Geophys. Res. Space Phys. 102 (A2), 2309–2324.

Blake, J., Carranza, P., Claudepierre, S., Clemmons, J., Crain, W., Dotan, Y., Fennell, J., Fuentes, F., Galvan, R., George, J., et al., 2013. The magnetic electron ion spectrometer (mageis) instruments aboard the radiation belt storm probes (rbsp) spacecraft.

Boscher, D., Bourdarie, S., O'Brien, P., Guild, T., 2013. The international radiation belt environment modeling (irbem) library.

Boyd, A.J., Spence, H.E., Claudepierre, S., Fennell, J.F., Blake, J., Baker, D., Reeves, G., Turner, D., 2014. Quantifying the radiation belt seed population in the 17 march 2013 electron acceleration event. Geophys. Res. Lett. 41 (7), 2275–2281.

Boyle, C., Reiff, P., Hairston, M., 1997. Empirical polar cap potentials. J. Geophys. Res. Space Phys. 102 (A1), 111–125.

Brautigam, D., Albert, J., 2000. Radial diffusion analysis of outer radiation belt electrons during the october 9, 1990, magnetic storm. J. Geophys. Res. Space Phys. 105 (A1), 291–309.

Büchner, J., Zelenyi, L., 1987. Chaotization of the electron motion as the cause of an internal magnetotail instability and substorm onset. J. Geophys. Res. Space Phys. 92 (A12), 13456–13466.

Carpenter, D., Anderson, R., 1992. An isee/whistler model of equatorial electron density in the magnetosphere. J. Geophys. Res. Space Phys. 97 (A2), 1097–1108.

Chen, M.W., Lemon, C.L., Orlova, K., Shprits, Y., Hecht, J., Walterscheid, R., 2015. Comparison of simulated and observed trapped and precipitating electron fluxes during a magnetic storm. Geophys. Res. Lett. 42 (20), 8302–8311.

Craven, J.D., 1966. Temporal variations of electron intensities at low altitudes in the outer radiation zone as observed with satellite injun 3. J. Geophys. Res. 71 (23), 5643–5663.

Data Book GOES, N., 2005. Data Book, Prepared for National Aeronautics and Space Administration Goddard Space Flight Center Greenbelt, Maryland 20771. Tech. Rep. CDRL PM-1-1-03, Section 5, pp. 5–6.

Delcourt, D., Sauvaud, J.-A., Martin, R., Moore, T., 1996. On the nonadiabatic precipitation of ions from the near-earth plasma sheet. J. Geophys. Res. Space Phys. 101 (A8), 17409–17418.

Denton, R., Takahashi, K., Galkin, I., Nsumei, P., Huang, X., Reinisch, B., Anderson, R., Sleeper, M., Hughes, W., 2006. Distribution of density along magnetospheric field lines. J. Geophys. Res. Space Phys. 111 (A4).

Dichter, B., McGarity, J., Oberhardt, M., Jordanov, V., Sperry, D., Huber, A., Pantazis, J., Mullen, E., Ginet, G., Gussenhoven, M., 1998. Compact environmental anomaly sensor (cease): A novel spacecraft instrument for in situ measurements of environmental conditions. IEEE Trans. Nucl. Sci. 45 (6), 2758–2764.

Drozdov, A., Shprits, Y., Aseev, N., Kellerman, A., Reeves, G.D., 2017. Dependence of radiation belt simulations to assumed radial diffusion rates tested for two empirical models of radial transport. Space Weather 15 (1), 150–162.

Drozdov, A., Shprits, Y., Orlova, K., Kellerman, A., Subbotin, D., Baker, D., Spence, H.E., Reeves, G., 2015. Energetic, relativistic, and ultrarelativistic electrons: Comparison of long-term verb code simulations with Van Allen probes measurements. J. Geophys. Res. Space Phys. 120 (5), 3574–3587.

Dubyagin, S., Ganushkina, N.Y., Sillanpää, I., Runov, A., 2016. Solar wind-driven variations of electron plasma sheet densities and temperatures beyond geostationary orbit during storm times. J. Geophys. Res. Space Phys. 121 (9), 8343–8360.

Elkington, S.R., Wiltberger, M., Chan, A.A., Baker, D.N., 2004. Physical models of the geospace radiation environment. J. Atmos. Solar-Terrestrial Phys. 66 (15), 1371–1387.

Fei, Y., Chan, A.A., Elkington, S.R., Wiltberger, M.J., 2006. Radial diffusion and mhd particle simulations of relativistic electron transport by ulf waves in the september 1998 storm. J. Geophys. Res. Space Phys. 111 (A12).

Fok, M.-C., Glocer, A., Zheng, Q., Horne, R.B., Meredith, N.P., Albert, J., Nagai, T., 2011. Recent developments in the radiation belt environment model. J. Atmos. Sol.-Terr. Phys. 73 (11–12), 1435–1443.

Fok, M.-C., Horne, R.B., Meredith, N.P., Glauert, S.A., 2008. Radiation belt environment model: Application to space weather nowcasting. J. Geophys. Res. Space Phys. 113 (A3).

Fu, H.S., Khotyaintsev, Y.V., André, M., Vaivads, A., 2011. Fermi and betatron acceleration of suprathermal electrons behind dipolarization fronts. Geophys. Res. Lett. 38 (16).

Ganushkina, N.Y., Amariutei, O., Shprits, Y., Liemohn, M., 2013. Transport of the plasma sheet electrons to the geostationary distances. J. Geophys. Res. Space Phys. 118 (1), 82–98.

Ganushkina, N.Y., Amariutei, O., Welling, D., Heynderickx, D., 2015. Nowcast model for low-energy electrons in the inner magnetosphere. Space Weather 13 (1), 16–34.

Ganushkina, N.Y., Liemohn, M., Amariutei, O., Pitchford, D., 2014. Low-energy electrons (5–50 kev) in the inner magnetosphere. J. Geophys. Res. Space Phys. 119 (1), 246–259.

Ganushkina, N.Y., Pulkkinen, T., Fritz, T., 2005. Role of substorm-associated impulsive electric fields in the ring current development during storms. Ann. Geophys. 23, 579–591.

Glauert, S.A., Horne, R.B., 2005. Calculation of pitch angle and energy diffusion coefficients with the padie code. J. Geophys. Res. Space Phys. 110 (A4).

Glauert, S.A., Horne, R.B., Meredith, N.P., 2014. Three-dimensional electron radiation belt simulations using the bas radiation belt model with new diffusion models for chorus, plasmaspheric hiss, and lightning-generated whistlers. J. Geophys. Res. Space Phys. 119 (1), 268–289.

Haiducek, J.D., Welling, D.T., Ganushkina, N.Y., Morley, S.K., Ozturk, D.S., 2017. Swmf global magnetosphere simulations of january 2005: Geomagnetic indices and cross-polar cap potential. Space Weather 15 (12), 1567–1587.

Hanser, F., 2011. Eps/Hepad Calibration and Data Handbook. Tech. Rep. goesn-eng-048d, Assurance Technology Corporation, Carlisle, Ma.

Horne, R.B., Glauert, S.A., Meredith, N.P., Koskinen, H., Vainio, R., Afanasiev, A., Ganushkina, N.Y., Amariutei, O.A., Boscher, D., Sicard, A., et al., 2013. Forecasting the earth's radiation belts and modelling solar energetic particle events: Recent results from spacecast. J. Space Weather Space Clim. 3 (A20).

Horne, R.B., Thorne, R.M., 1998. Potential waves for relativistic electron scattering and stochastic acceleration during magnetic storms. Geophys. Res. Lett. 25 (15), 3011–3014.

Horne, R.B., Thorne, R.M., Glauert, S.A., Meredith, N.P., Pokhotelov, D., Santolík, O., 2007. Electron acceleration in the Van Allen radiation belts by fast magnetosonic waves. Geophys. Res. Lett. 34 (17).

Horne, R.B., Thorne, R.M., Shprits, Y.Y., Meredith, N.P., Glauert, S.A., Smith, A.J., Kanekal, S.G., Baker, D.N., Engebretson, M.J., Posch, J.L., et al., 2005. Wave acceleration of electrons in the Van Allen radiation belts. Nature 437 (7056), 227–230.

Hudson, M., Paral, J., Kress, B., Wiltberger, M., Baker, D., Foster, J., Turner, D., Wygant, J.R., 2015. Modeling cme-shock-driven storms in 2012–2013: Mhd test particle simulations. J. Geophys. Res. Space Phys. 120 (2), 1168–1181.

Jaynes, A., Baker, D., Singer, H., Rodriguez, J., Loto'aniu, T., Ali, A., Elkington, S., Li, X., Kanekal, S., Claudepierre, S., et al., 2015. Source and seed populations for relativistic electrons: Their roles in radiation belt changes. J. Geophys. Res. Space Phys. 120 (9), 7240–7254.

Jordanova, V., Albert, J., Miyoshi, Y., 2008. Relativistic electron precipitation by emic waves from self-consistent global simulations. J. Geophys. Res. Space Phys. 113 (A3).

Jordanova, V., Kozyra, J.-A., Nagy, A., Khazanov, G., 1997. Kinetic model of the ring current-atmosphere interactions. J. Geophys. Res. Space Phys. 102 (A7), 14279–14291.

Jordanova, V., Miyoshi, Y., 2005. Relativistic model of ring current and radiation belt ions and electrons: Initial results. Geophys. Res. Lett. 32 (14).

Jordanova, V., Welling, D., Zaharia, S., Chen, L., Thorne, R., 2012. Modeling ring current ion and electron dynamics and plasma instabilities during a high-speed stream driven storm. J. Geophys. Res. Space Phys. 117 (A9).

Kennel, C., Engelmann, F., 1966. Velocity space diffusion from weak plasma turbulence in a magnetic field. Phys. Fluids 9 (12), 2377–2388.

Kim, K.-C., Shprits, Y., Subbotin, D., Ni, B., 2011. Understanding the dynamic evolution of the relativistic electron slot region including radial and pitch angle diffusion. J. Geophys. Res. Space Phys. 116 (A10).

Lanzerotti, L.J., 2001. Space weather effects on technologies. Space weather 11–22.

Lenchek, A., Singer, S., Wentworth, R., 1961. Geomagnetically trapped electrons from cosmic ray albedo neutrons. J. Geophys. Res. 66 (12), 4027–4046.

Lerche, I., 1968. Quasilinear theory of resonant diffusion in a magneto-active, relativistic plasma. Phys. Fluids 11 (12), 1720–1727.

Li, X., Baker, D., Temerin, M., Reeves, G., Belian, R., 1998. Simulation of dispersionless injections and drift echoes of energetic electrons associated with substorms. Geophys. Res. Lett. 25 (20), 3763–3766.

Li, Z., Hudson, M., Patel, M., Wiltberger, M., Boyd, A., Turner, D., 2017. Ulf wave analysis and radial diffusion calculation using a global MHD model for the 17 march 2013 and 2015 storms. J. Geophys. Res. Space Phys. 122 (7), 7353–7363.

Li, W., Ma, Q., Thorne, R., Bortnik, J., Zhang, X.J., Li, J., Baker, D., Reeves, G.D., Spence, H., Kletzing, C., et al., 2016. Radiation belt electron acceleration during the 17 march 2015 geomagnetic storm: Observations and simulations. J. Geophys. Res. Space Phys. 121 (6), 5520–5536.

Li, W., Shprits, Y., Thorne, R., 2007. Dynamic evolution of energetic outer zone electrons due to wave-particle interactions during storms. J. Geophys. Res. Space Phys. 112 (A10).

Li, W., Thorne, R., Bortnik, J., Nishimura, Y., Angelopoulos, V., Chen, L., McFadden, J., Bonnell, J., 2010. Global distributions of suprathermal electrons observed on themis and potential mechanisms for access into the plasmasphere. J. Geophys. Res. Space Phys. 115 (A12).

Li, W., Thorne, R., Ma, Q., Ni, B., Bortnik, J., Baker, D., Spence, H.E., Reeves, G., Kanekal, S., Green, J., et al., 2014. Radiation belt electron acceleration by chorus waves during the 17 march 2013 storm. J. Geophys. Res. Space Phys. 119 (6), 4681–4693.

Liemohn, M., Ganushkina, N.Y., D. Zeeuw, D.L., Rastaetter, L., Kuznetsova, M., Welling, D.T., Toth, G., Ilie, R., Gombosi, T.I., van der Holst, B., 2018. Real-time swmf at ccmc: Assessing the dst output from continuous operational simulations. Space Weather 16 (10), 1583–1603.

Liu, S., Chen, M., Lyons, L., Korth, H., Albert, J., Roeder, J., Anderson, P., Thomsen, M., 2003. Contribution of convective transport to stormtime ring current electron injection. J. Geophys. Res. Space Phys. 108 (A10).

Liu, S., Chen, M., Roeder, J., Lyons, L., Schulz, M., 2005. Relative contribution of electrons to the stormtime total ring current energy content. Geophys. Res. Lett. 32 (3).

Lyons, L., 1984. Electron energization in the geomagnetic tail current sheet. J. Geophys. Res. Space Phys. 89 (A7), 5479–5487.

Lyons, L., Thorne, R., Kennel, C., 1971. Electron pitch-angle diffusion driven by oblique whistler-mode turbulence. J. Plasma Phys. 6 (3), 589–606.

Mauk, B., Fox, N.J., Kanekal, S., Kessel, R., Sibeck, D., Ukhorskiy, A., 2012. Science objectives and rationale for the radiation belt storm probes mission. In: The Van Allen Probes Mission. Springer, pp. 3–27.

Meredith, N.P., Horne, R.B., Glauert, S.A., Anderson, R.R., 2007. Slot region electron loss timescales due to plasmaspheric hiss and lightning-generated whistlers. J. Geophys. Res. Space Phys. 112 (A8).

Millan, R., Baker, D., 2012. Acceleration of particles to high energies in earth's radiation belts. Space Sci. Rev. 173 (1–4), 103–131.

Miyoshi, Y., Jordanova, V., Morioka, A., Thomsen, M., Reeves, G., Evans, D., Green, J., 2006. Observations and modeling of energetic electron dynamics during the october 2001 storm. J. Geophys. Res. Space Phys. 111 (A11).

Miyoshi, Y., Morioka, A., Misawa, H., Obara, T., Nagai, T., Kasahara, Y., 2003. Rebuilding process of the outer radiation belt during the 3 november 1993 magnetic storm: Noaa and exos-d observations. J. Geophys. Res. Space Phys. 108 (A1).

Ni, B., Thorne, R.M., Shprits, Y.Y., Bortnik, J., 2008. Resonant scattering of plasma sheet electrons by whistler-mode chorus: Contribution to diffuse auroral precipitation. Geophys. Res. Lett. 35 (11).

Odenwald, S., Green, J., Taylor, W., 2006. Forecasting the impact of an 1859-calibre superstorm on satellite resources. Adv. Space Res. 38 (2), 280–297.

Orlova, K., Shprits, Y., 2014. Model of lifetimes of the outer radiation belt electrons in a realistic magnetic field using realistic chorus wave parameters. J. Geophys. Res. Space Phys. 119 (2), 770–780.

Orlova, K., Shprits, Y., Spasojevic, M., 2016. New global loss model of energetic and relativistic electrons based on Van Allen probes measurements. J. Geophys. Res. Space Phys. 121, 1308–1314.

Orlova, K., Spasojevic, M., Shprits, Y., 2014. Activity-dependent global model of electron loss inside the plasmasphere. Geophys. Res. Lett. 41 (11), 3744–3751.

Ozeke, L.G., Mann, I.R., Murphy, K.R., Jonatha. Rae, I., Milling, D.K., 2014. Analytic expressions for ulf wave radiation belt radial diffusion coefficients. J. Geophys. Res. Space Phys. 119 (3), 1587–1605.

Parsons, A., Biesecker, D., Odstrcil, D., Millward, G., Hill, S., Pizzo, V., 2011. Wang-Sheeley-Arge–Enlil cone model transitions to operations. Space Weather 9 (3).

Reeves, G., Baker, D., Belian, R., Blake, J., Cayton, T., Fennell, J., Friedel, R., Meier, M., Selesnick, R., Spence, H.E., 1998. The global response of relativistic radiation belt electrons to the january 1997 magnetic cloud. Geophys. Res. Lett. 25 (17), 3265–3268.

Ripoll, J.-F., Santolík, G., Kurth, W., Denton, M., Loridan, V., Thaller, S., Kletzing, C., Turner, D., 2017. Effects of whistler mode hiss waves in march 2013. J. Geophys. Res. Space Phys. 122 (7), 7433–7462.

Rodriguez, J., 2014a. Goes 13–15 Mage/Pd Pitch Angles Algorithm Theoretical Basis Document, Version 1.0. Noaa National Geophysical Data Center, Boulder, CO.

Rodriguez, J., 2014b. Goes Epead Science-Quality Electron Fluxes Algorithm Theoretical Basis Document. NOAA Nat. Geophys. Data Center.

Rothwell, P., McIlwain, C.E., 1960. Magnetic storms and the Van Allen radiation belts—observations from satellite 1958ε (explorer iv). J. Geophys. Res. 65 (3), 799–806.

Sarris, T.E., Li, X., Tsaggas, N., Paschalidis, N., 2002. Modeling energetic particle injections in dynamic pulse fields with varying propagation speeds. J. Geophys. Res. Space Phys. 107 (A3).

Schulz, M., Lanzerotti, L.J., 1974. Particle diffusion in the radiation belts. Phys. Chem. Space 7.

Sergeev, V., Tsyganenko, N., 1982. Energetic particle losses and trapping boundaries as deduced from calculations with a realistic magnetic field model. Planet. Space Sci. 30 (10), 999–1006.

Sheeley, B., Moldwin, M., Rassoul, H., Anderson, R., 2001. An empirical plasmasphere and trough density model: Crres observations. J. Geophys. Res. Space Phys. 106 (A11), 25631–25641.

Shi, R., Summers, D., Ni, B., Fennell, J.F., Blake, J.B., Spence, H.E., Reeves, G.D., 2016. Survey of radiation belt energetic electron pitch angle distributions based on the Van Allen probes mageis measurements. J. Geophys. Res. Space Phys. 121 (2), 1078–1090.

Shprits, Y.Y., Chen, L., Thorne, R.M., 2009a. Simulations of pitch angle scattering of relativistic electrons with mlt-dependent diffusion coefficients. J. Geophys. Res. Space Phys. 114 (A3).

Shprits, Y.Y., Elkington, S.R., Meredith, N.P., Subbotin, D.A., 2008a. Review of modeling of losses and sources of relativistic electrons in the outer radiation belt I: Radial transport. J. Atmos. Sol.-Terr. Phys. 70 (14), 1679–1693.

Shprits, Y.Y., Kellerman, A.C., Drozdov, A.Y., Spence, H.E., Reeves, G.D., Baker, D.N., 2015. Combined convective and diffusive simulations: Verb-4d comparison with 17 march 2013 Van Allen probes observations. Geophys. Res. Lett. 42 (22), 9600–9608.

Shprits, Y., Kondrashov, D., Chen, Y., Thorne, R., Ghil, M., Friedel, R., Reeves, G., 2007a. Reanalysis of relativistic radiation belt electron fluxes using crres satellite data, a radial diffusion model, and a kalman filter. J. Geophys. Res. Space Phys. 112 (A12).

Shprits, Y.Y., Meredith, N.P., Thorne, R.M., 2007b. Parameterization of radiation belt electron loss timescales due to interactions with chorus waves. Geophys. Res. Lett. 34 (11).

Shprits, Y.Y., Ni, B., 2009. Dependence of the quasi-linear scattering rates on the wave normal distribution of chorus waves. J. Geophys. Res. Space Phys. 114 (A11).

Shprits, Y.Y., Subbotin, D.A., Meredith, N.P., Elkington, S.R., 2008b. Review of modeling of losses and sources of relativistic electrons in the outer radiation belt II: Local acceleration and loss. Journal of atmospheric and solar-terrestrial physics 70 (14), 1694–1713.

Shprits, Y.Y., Subbotin, D., Ni, B., 2009b. Evolution of electron fluxes in the outer radiation belt computed with the verb code. J. Geophys. Res. Space Phys. 114 (A11).

Shprits, Y., Thorne, R., Friedel, R., Reeves, G., Fennell, J., Baker, D., Kanekal, S., 2006a. Outward radial diffusion driven by losses at magnetopause. J. Geophys. Res. Space Phys. 111 (A11).

Shprits, Y., Thorne, R., Horne, R., Glauert, S., Cartwright, M., Russell, C., Baker, D., Kanekal, S., 2006b. Acceleration mechanism responsible for the formation of the new radiation belt during the 2003 halloween solar storm. Geophys. Res. Lett. 33 (5).

Shprits, Y., Thorne, R., Reeves, G., Friedel, R., 2005. Radial diffusion modeling with empirical lifetimes: Comparison with crres observations. Ann. Geophys. 1467–1471.

Starks, M., Quinn, R., Ginet, G., Albert, J., Sales, G., Reinisch, B., Song, P., 2008. Illumination of the plasmasphere by terrestrial very low frequency transmitters: Model validation. J. Geophys. Res. Space Phys. 113 (A9).

Su, Z., Zhu, H., Xiao, F., Zong, Q.G., Zhou, X.Z., Zheng, H., Wang, Y., Wang, S., Hao, Y.X., Gao, Z., et al., 2015. Ultra-low-frequency wave-driven diffusion of radiation belt relativistic electrons. Nature Commun. 6, 10096.

Subbotin, D., Shprits, Y., 2009. Three-dimensional modeling of the radiation belts using the versatile electron radiation belt (verb) code. Space Weather 7 (10).

Subbotin, D., Shprits, Y., 2012. Three-dimensional radiation belt simulations in terms of adiabatic invariants using a single numerical grid. J. Geophys. Res. Space Phys. 117 (A5).

Subbotin, D., Shprits, Y., Gkioulidou, M., Lyons, L., Ni, B., Merkin, V., Toffoletto, F., Thorne, R., Horne, R.B., Hudson, M., 2011a. Simulation of the acceleration of relativistic electrons in the inner magnetosphere using rcm-verb coupled codes. J. Geophys. Res. Space Phys. 116 (A8).

Subbotin, D., Shprits, Y., Ni, B., 2010. Three-dimensional verb radiation belt simulations including mixed diffusion. J. Geophys. Res. Space Phys. 115 (A3).

Subbotin, D., Shprits, Y., Ni, B., 2011b. Long-term radiation belt simulation with the verb 3-d code: Comparison with crres observations. J. Geophys. Res. Space Phys. 116 (A12).

Summers, D., Thorne, R.M., Xiao, F., 1998. Relativistic theory of wave-particle resonant diffusion with application to electron acceleration in the magnetosphere. J. Geophys. Res. Space Phys. 103 (A9), 20487–20500.

Thorne, R., Li, W., Ni, B., Ma, Q., Bortnik, J., Baker, D., Spence, H.E., Reeves, G., Henderson, M., Kletzing, C., et al., 2013. Evolution and slow decay of an unusual narrow ring of relativistic electrons near l∼ 3.2 following the september 2012 magnetic storm. Geophys. Res. Lett. 40 (14), 3507–3511.

Toffoletto, F., Sazykin, S., Spiro, R., Wolf, R., 2003. Inner magnetospheric modeling with the rice convection model. Space Sci. Rev. 107 (1–2), 175–196.

Tsyganenko, N.A., 1995. Modeling the earth's magnetospheric magnetic field confined within a realistic magnetopause. J. Geophys. Res. Space Phys. 100 (A4), 5599–5612.

Tsyganenko, N., Sitnov, M., 2005. Modeling the dynamics of the inner magnetosphere during strong geomagnetic storms. J. Geophys. Res. Space Phys. 110 (A3).

Tu, W., Cunningham, G., Chen, Y., Morley, S., Reeves, G., Blake, J., Baker, D., Spence, H., 2014. Event-specific chorus wave and electron seed population models in dream3d using the Van Allen probes. Geophys. Res. Lett. 41 (5), 1359–1366.

Varotsou, A., Boscher, D., Bourdarie, S., Horne, R.B., Glauert, S.A., Meredith, N.P., 2005. Simulation of the outer radiation belt electrons near geosynchronous orbit including both radial diffusion and resonant interaction with whistler-mode chorus waves. Geophys. Res. Lett. 32 (19).

Varotsou, A., Boscher, D., Bourdarie, S., Horne, R.B., Meredith, N.P., Glauert, S.A., Friedel, R.H., 2008. Simulation of the outer radiation belt electron dynamics including electron-chorus resonant interactions. J. Geophys. Res. Space Phys. 113 (A12).

Walt, M., 1994. Introduction to geomagnetically trapped radiation. Camb. Atmos. Space Sci. Ser. 10, 10.

Wintoft, P., Wik, M., 2018. Evaluation of kp and dst predictions using ace and dscovr solar wind data. Space Weather 16 (12), 1972–1983.

Yu, Y., Jordanova, V., Welling, D., Larsen, B., Claudepierre, S.G., Kletzing, C., 2014. The role of ring current particle injections: Global simulations and Van Allen probes observations during 17 march 2013 storm. Geophys. Res. Lett. 41 (4), 1126–1132.

Zaharia, S., Jordanova, V., Welling, D., Tóth, G., 2010. Self-consistent inner magnetosphere simulation driven by a global mhd model. J. Geophys. Res. Space Phys. 115 (A12).